\documentclass[]{fcs}  % Frontiers in Computer Science (fcs) class
\usepackage{graphicx}  % For including graphics
\usepackage{tcolorbox} % For the custom prompt box
\usepackage{listings}  % For code listings
\usepackage{amsmath}   % Optional, for equations
\usepackage{lipsum}    % For filler text (optional)
\usepackage{multicol}  % To handle multiple columns
\usepackage{multirow}

\newtcolorbox{promptbox}[1][]{
    colback=gray!5,           % Very light gray background (5% gray)
    colframe=gray!80!black,   % Dark gray frame (80% gray + black)
    coltitle=white,           % White title color for contrast
    fonttitle=\bfseries,      % Font style for the title
    title=#1,                 % Title text
    boxrule=0.8mm,            % Thickness of the frame
    sharp corners,            % Sharp corners for the box
    width=\textwidth           % Full text width for use in table*
}

\title{ An Improved Quantum Software Challenges Classification Approach using Transfer Learning and Explainable AI }
\author*[1]{Nek Dil Khan}
\author*[2]{Javed Ali Khan}
\author*[3]{Mobashir Husain}
\author*[3]{Muhammad Sohail Khan}
\author*[4]{Arif Ali Khan}
\author*[5]{Muhammad Azeem Akbar}
\author*[6]{Shahid Hussain}

\address[1]{Faculty of Information Technology, Beijing University of Technology, Beijing 100081, China}
\address[2]{Department of Computer Science, University of Hertfordshire, College Lane, Hatfield, AL10 9AB, United Kingdom}
\address[3]{Department of Computer Software Engineering, University of Engineering and Technology, Mardan, Pakistan}
\address[4]{M3S Empirical Software Engineering Research Unit, University of Oulu, Oulu, Finland}
\address[5]{Department of Software Engineering, LUT University, Lappeenranta, Finland}
\address[6]{Department of Computer Science and Software Engineering, Penn State University Behrend, 242 Burke Center, Erie, PA, 16563, USA}

\fcssetup{
  corr-email     = {j.a.khan@herts.ac.uk}, % ← Corresponding author email
}
\begin{abstract}
 Quantum Software Engineering (QSE) has emerged as a research direction practiced by tech joints. However, quantum developers face challenges in optimizing quantum computing and QSE concepts. Quantum developers use Stack Overflow (SO) to discuss quantum-related challenges and use specialized quantum tags to label posts in developers' forums. These tags often refer to technical quantum aspects rather than developer posts. Categorizing quantum practitioners' questions based on quantum concepts can help developers identify frequent QSE challenges. We conducted qualitative and quantitative studies to classify quantum developers' questions into various challenges. We extracted 2829 developers' questions from Q\&A platforms using quantum-related tags. The posts were analyzed to identify frequent quantum-related challenges and develop a novel grounded theory. The challenges identified include Tooling, Theoretical, Learning, Conceptual, Errors, and API Usage. Through content analysis and grounded theory, the developers' discussions were annotated with commonly reported quantum challenges to develop a ground truth dataset. ChatGPT was used to validate human annotations and resolve disagreements. Various fine-tuned transformer algorithms, including BERT, DistilBERT, and RoBERTa, were used to classify developer discussions into commonly reported quantum challenges. We achieved an average accuracy of 95\% with BERT DistilBERT algorithms, compared to fine-tuned Deep and Machine Learning (D\&ML) classifiers, including Feedforward Neural Networks (FNN), Convolutional Neural Networks (CNN), and Long Short-Term Memory networks (LSTM), which achieved accuracies of 89\%, 86\%, and 84\%, respectively. The proposed Transformer-based approach outperforms the previous D\&ML-based approach with a 6\% increase in accuracy by processing actual developer discussions, i.e., without data augmentation. Furthermore, we applied SHAP (SHapley Additive exPlanations) to provide model interpretability, revealing how specific linguistic features drive predictions and enhancing transparency in the classification process. These improved research findings can help quantum vendors and developers' discussion forums to better organize developers' discussions for improved access and readability. However, it further needs empirical evaluation studies with actual quantum developers and vendors. 
 
\end{abstract}

\begin{document}
\section{Introduction}
Quantum computing leverages quantum mechanics principles to describe the behavior of particles at the atomic and subatomic levels. Unlike classical computers, which use bits (0s and 1s), quantum computers use quantum bits or "qubits" to process information \cite{nielsen2010quantum}. Superposition allows qubits to exist in a combination of 0 and 1 states simultaneously, increasing the computational power of quantum systems \cite{preskill1998reliable}. Google \cite{arute2019quantum}, IBM \cite{ibm2023quantum}, and Rigetti \cite{rigetti2020why} have demonstrated the ability to create qubits in controlled environments, enabling practical quantum computing. This provides opportunities to solve complex computational problems requiring parallel computation, which is challenging for classical computers. It can resolve complex issues in big data analytics, financial systems, national security, chemistry, cryptography, health analytics, and medicine \cite{piattini2020talavera}.

Recently, there has been an increasing interest from the research community in proposing various approaches across different QSE activities \cite{murillo2025quantum}. Where QSE is a discipline emphasizing designing, developing, and optimizing software for a wide range of applications \cite{mandal2025quantum}. QSE processes must be aligned with hardware advancements to leverage the capabilities of quantum computing effectively. Developing methodologies, processes, tools, and algorithms is crucial for quantum computing domains \cite{zhao2020quantum}. Quantum programming languages like Q\# \cite{svore2018q}, Scaffold \cite{javadiabhari2015scaffcc}, and Qiskit \cite{qiskit2024} exist for developing quantum applications. Microsoft, Google, and IBM have developed cloud platforms that support the development of quantum software. IBM's platform enables the execution of applications on quantum systems. Developers create quantum software for domains such as chemistry and machine learning \cite{li2021understanding}. Given the importance of applications, software engineering processes are required for large projects. The QSE was coined to meet the demands for large applications \cite{zhao2020quantum}. Initial approaches for quantum requirements engineering \cite{yue2023towards}, quantum software architecture \cite{ahmad2022towards}, Quantum programming \cite{khan2025mining},  Quantum Development \cite{akbar2023systematic}, and Quantum Testing \cite{ali2023quantum} are emerging. Software practitioners use developers' discussion forums to extract information and address challenges associated with software evolution \cite{ahmad2018survey} \cite{de2025oss}. For example, researchers have analyzed Q\&A platform discussions to identify challenges related to software development methodologies \cite{khan2024insights}. Also, De Dieu et al. \cite{de2025oss} explore Q\&A for extracting software architecture related information. Similarly, Li et al. \cite{li2024machine} explore Q\&A for extracting requirements-related information. Moreover, El-Aoun used Q\&A to identify frequently occurring challenges to the quantum developers community \cite{li2021understanding}. These forums offer resources, communities, and opportunities for problem-solving. Q\&A are valuable for sharing experiences in less-resourced areas, such as QSE, blockchain, etc. They enable collaboration, updates on trends, and collective problem-solving. The developers' posts in these Q\&A platforms are categorized based on specific tags, which we believe they hard and challenging for developers to remember. Particularly, the more emerging and comparatively new domains, i.e., QSE. Therefore, this motivated us to contextually analyze the developers' questions to identify their best possible category and add it to the existing tags associated with it from improved searching and visibility in the Q\&A forums, which is aligned with previous research \cite{beyer2020kind}. 

In this study, we evaluate developer discussions on QSE across various Q\&A platforms like Stack Overflow (SO), Quantum Computing Stack Exchange (QCSE), Computer Science Stack Exchange (CSSE), and Artificial Intelligence Stack Exchange (AISE) to understand their contextual category, which are often reported as challenges by software quantum practitioners. This aids quantum developers in finding possible solutions to specific QSE-related challenges on these platforms that are reported by other quantum practitioners and are frequently organized by tags such as qiskit, qcl, qutip, qubit, and TensorFlow Quantum. For this purpose,  we extracted 2829 developers' questions using quantum-related tags, following El-Aoun et al.'s method \cite{li2021understanding}. These developers' posts were manually critically analyzed to identify frequently discussed quantum challenges and develop a novel grounded theory, which was cross-validated with that of El-Aoun et al. \cite{li2021understanding} who identified QSE challenges using topic modeling. The frequent challenges include Tooling, Theoretical, Learning, Conceptual, Errors, and API Usage. Next, using content analysis and the novel QSE grounded theory document, the developers' discussions in the dataset were annotated with common quantum challenges to create a ground-truth dataset for the D\&ML and transfer learning (TL) algorithms. Moreover, to validate the annotation process, ChatGPT was utilized to annotate developers' discussions. Additionally, another cycle was run with ChatGPT to resolve disagreements between the human annotators and ChatGPT, resulting in a conflict-free dataset. Finally, in this approach, in addition to the various fine-tuned D\&ML classifiers, fine-tuned transformer learning algorithms, including BERT, DistilBERT, and RoBerta, were used to contextually analyze developer discussions and classify them into common quantum challenges.   

We extended the work published in the Proceedings of the 33rd ACM International Conference on the Foundations of Software Engineering \cite{husain2025exploring}, with the following additions.
\begin{itemize}

\item In addition to the D\&ML experiments, we aim to propose a more generalised and robust approach by employing fine-tuned transformer-learning algorithms, achieving comparatively better performance in classifying developers' feedback into fine-grained quantum types by utilising transformers' contextual embeddings and pretrained knowledge.
\item We extended the introduction by adding more recent and relevant QSE references and motivation for the proposed approach.
\item We critically analysed developers' feedback using a content analysis approach to identify the frequency of quantum-related concepts in the Q\&A forums.
\item We extended the QSE dataset annotation using ChatGPT by adding various prompts to improve the understanding of complex annotation processes.
\item We fine-tuned state-of-the-art transformer algorithms for improved accuracy and compared their performances with fine-tuned D\&ML algorithms. Additionally, various confusion matrices are developed to demonstrate the performance of fine-tuned transformers and D\&ML algorithms in achieving lower rates of false positives and negatives. 
\item We used SHAP as an explainability layer for TL models in the QSE domain, providing global and local interpretability to reveal how specific linguistic features influence model predictions, thereby enhancing transparency and trust in automated classification systems.
\item We extended and refined the discussion of the manuscript with more insightful implications of the proposed approach for quantum developers and software vendors. 

\end{itemize}

The paper is organized as follows: Section 2 provides a related work. Section 3 outlines the research methodology. Section 4 details the classification and analysis of challenges faced in QSE. Section 5 explores automated methods for classifying developer discussions. Section 6 discusses the insights derived from developer exchanges on platforms like Stack Exchange. Section 7 concludes the paper and discusses possible future directions.

\section{RELATED WORK}
This section reviews the pertinent literature on quantum computing, QSE, and mining software repositories. 
\subsection{Software Quantum-Based Approaches}
This subsection reviews the literature on software quantum computing approaches and techniques for quantum software development. Gill et al.\cite{gill2022quantum} provided a systematic review of Quantum Computing (QC), which offers computational advantages over classical computing by leveraging quantum mechanical principles. They highlighted QC's potential to address complex problems in drug design, data science, clean energy, finance, industrial chemical development, secure communications, and quantum chemistry. Akbar et al. \cite{akbar2023systematic} proposed a perspective on QSE, outlining its lifecycle stages and providing a framework for quantum necessity engineering, software implementation, design, testing, and maintenance. El-Aoun et al.\cite{li2021understanding} explored challenging aspects of QSE, examining quantum code theory and the gap between quantum and classical computing. Vietz et al. \cite{vietz2021decision} analyzed tools supporting quantum application development, providing categorization and analysis to aid developers in selecting suitable tools. Haghparast et al.\cite{haghparast2023quantum} explored QSE challenges, mapped research challenges to the quantum computing workflow model, and identified directions for software engineering research.
\subsection{Mining Q\&A repositors}
We elaborate on the approaches that mine Q\&A repositories to enhance developers' understanding. Beyer et al. \cite{beyer2020kind} proposed a method to classify developer questions in SO forums into seven categories. They gathered data from SO forums on Android developers' questions and classified them using ML classifiers and expressions. Treude et al. \cite{treude2011programmers} studied how programmers ask and answer questions on SO, analyzing discussions to categorize question types and assess answers. Their findings show the utility of Q\&A sites for code reviews and conceptual questions. Iftikhar et al.\cite{iftikhar2021deep} proposed a deep learning approach for predicting correct answers in SO discussions. They extracted metadata and question/answer combinations, applied NLP techniques, and used keyword ranking for feature vectors. Their ensemble deep learning model surpassed state-of-the-art methods, improving the accuracy, precision, recall, and f-measure by 1.72\%, 24.96\%, 6.57\%, and 16.62\%, respectively. Zhu et al. \cite{zhu2022empirical} studied SO discussions, noting active participation in Q\&A. They found a strong link between question comments and the response time. Similarly, \cite{khan2022valuating} analyzed Q\&A forum discussions to gain insights into software development methods. El-Aoun et al.\cite{li2021understanding} analyzed Q\&A platforms to uncover challenges faced by quantum developers. Khan et al. \cite{ali2020requirements} proposed an automated approach for extracting requirement-related information from forums, using ML and NLP to categorize discussions. Beyer et al. manually categorized 450 SO posts on Android app development \cite{beyer2014manual}. They found 'How to?' and 'What is the problem?' were common, with issues related to the User Interface and core elements. Their study revealed correlations between the problem and question types.
\subsection{Comparison with Existing Literature}
The study complements mining software repositories and analyzing developer discussions to improve practitioner understanding. Inspired by El-Aoun et al. and Beyer et al. \cite{beyer2020kind}, our approach differs from previous studies. El-Aoun et al.\cite{li2021understanding} used topic modelling on Q\&A forums to identify challenges, while Bayer et al. \cite{beyer2020kind} classified developers' comments related to Android development into fine-grained categories. We critically analyzed QSE discussions from multiple forums to develop a taxonomy, validating it against El-Aoun et al.\cite{li2021understanding}'s findings. In our previously published paper on QSE \cite{husain2025exploring}, we fine-tuned ML and DL classifiers to classify QSE discussions into frequently occurring challenges. However, due to the comparatively emerging domain of QSE, the performance of D\&ML algorithms was relatively low due to limited data availability in the discussion forum. Still, to counter this, we employed a data augmentation approach to improve the D\&ML performances at the cost of adding additional complexity and risk of training the algorithms on synthetically generated data. In the proposed approach, we leverage the power of TL algorithms by fine-tuning them for the QSE problem to better understand the context of developers' discussions and classify them into more closely related QSE challenges.  Moreover, we employed existing explainability approaches with the fine-tuned transformer to enhance the interpretability of the transformer algorithms to understand the features frequently used by TL algorithms when deciding on a particular QSE challenge type. Additionally, unlike the Beyer et al. \cite{beyer2020kind} approach, we aimed for a generalizable approach using multiple QSE forums. We employed ChatGPT to validate coder disagreements and automate annotation. Table \ref{tab:devquestionclassification} compares existing approaches.

\begin{table*}[!tb]
\centering
\footnotesize
\renewcommand{\arraystretch}{1.3} % Increased row height for clarity
\caption{Summary of Methodologies for Automated Developer Question Classification}
\setlength{\tabcolsep}{5pt} % Adjusted spacing between columns
\begin{tabular}{p{2.0cm}p{2.0cm}p{5.0cm}p{2.0cm}p{3.5cm}c}
\hline
\textbf{Ref.} & \textbf{Algorithms Used} & \textbf{Proposed Methodology} & \textbf{Dataset Used and Size} & \textbf{Key Contributions} & \textbf{QC/QSE?} \\
\hline

Beyer et al. \cite{beyer2020kind} & 
RF, SVM & 
Automated classification via taxonomy harmonization (7 categories), regex/ML. & 
1,000 Android SO posts & 
Improved question search/browsing for developers. & 
No \\

Vietz et al. \cite{vietz2021decision} & 
- & 
Quantum dev tools taxonomy + comparison framework. & 
- & 
Structured quantum tech selection for developers. & 
Yes \\

Khan et al. \cite{khan2024insights} & 
LDA & 
Mixed-methods analysis (13,903 posts) with topic modeling. & 
13,903 SO/SESE/PMSE posts & 
Identified common dev approaches/challenges. & 
No \\

El Aoun et al. \cite{li2021understanding} & 
Topic Modeling & 
QSE challenge identification via Stack Exchange/GitHub. & 
Stack Exchange + GitHub & 
Highlighted QSE-specific challenges. & 
Yes \\

\textbf{Proposed Work} & 
\textbf{FNN, CNN, LSTM, GRU, RNN, BERT, RoBERa, DistilBERT} & 
\textbf{Classification of 2,829 quantum dev questions + ChatGPT validation + data augmentation.} & 
2,829 Q\&A posts & 
QSE challenge taxonomy + LLM validation + enhanced classification. & 
\textbf{Yes} \\
\hline

\end{tabular}
\label{tab:devquestionclassification}
\end{table*}

\section{PROPOSED RESEARCH METHODOLOGY}
In this section, we first elaborate on the research questions to answer the proposed research methodology. Next, we describe the proposed research approach, which includes the research dataset, novel grounded theory, content analysis, and various classifiers used to classify developer discussions into frequently reported challenge types. 
\subsection{Research Questions}
This study aims to develop an automated methodology for systematically classifying developer discussions into identified challenges using fine-tuning D\&ML and transfer algorithms. The proposed approach will empower QSE developers with structured resources, enabling them to find relevant information about the quantum challenges they face. We developed the following research questions to validate the proposed methodology:

\textbf{RQ1: What do quantum developers discuss about various challenges in different developers' discussion forums?}

\textbf{RQ2: What frequent QSE challenges can be identified from the developer's discussion?}

\textbf{RQ3: Can ChatGPT work as an annotator and negotiator in developing a ground truth for D\&ML and transformer classifiers?}

\textbf{RQ4: Do fine-tuned transformer algorithms outperform fine-tuned D\&ML algorithms in classifying actual developers' discussions into various QSE challenges?}

\textbf{RQ5: Does the SHAP explainable approach enhance the interpretability of fine-tuned transformer models in understanding classification decision-making?}

In pursuit of RQ-1, we aim to analyze developer questions in various forums, including SO, QCSE, CSSE, and AISE, to identify common patterns in how quantum developers express concerns or challenges regarding QSE. This will result in a novel grounded theory to identify frequently occurring challenges related to QSE. RQ2 utilizes the grounded theory and content analysis approach to develop a ground truth for RQ4 by manually annotating developers' discussions into various frequently occurring challenges. For RQ3, we seek to determine whether ChatGPT can annotate developers' discussions about QSE compared to human annotators and explore its use as a negotiator to resolve conflicts between annotators. This aims to overcome the manual complexity of annotating developer datasets for classification tasks. For RQ4, we aim to evaluate and compare various D\&ML and TL algorithms by exploring different feature engineering and hyperparameter tuning approaches to fine-tune these algorithms and identify their efficacy in classifying developers' discussions into SQE challenges identified through RQ1. Moreover, in the previously published work \cite{husain2025exploring}, the D\&ML algorithms revealed limitations in capturing the complex linguistic nuances of QSE-related developers' discussions by classifying them into frequently occuring QSE challenges due to limited data. In the proposed approach, we fine-tuned TL models, specifically BERT, DistilBERT, and RoBERTa, and evaluated their performance in classifying developers' discussions from Q\&A forums for QSE challenges, compared to D\&ML algorithms without data augmentation. Finally, RQ5 investigates the application of SHAP to these TL models to provide transparent and interpretable results.

\subsection{Research method}
\begin{figure*}[h]
  \centering
  \includegraphics[width=0.87\linewidth]{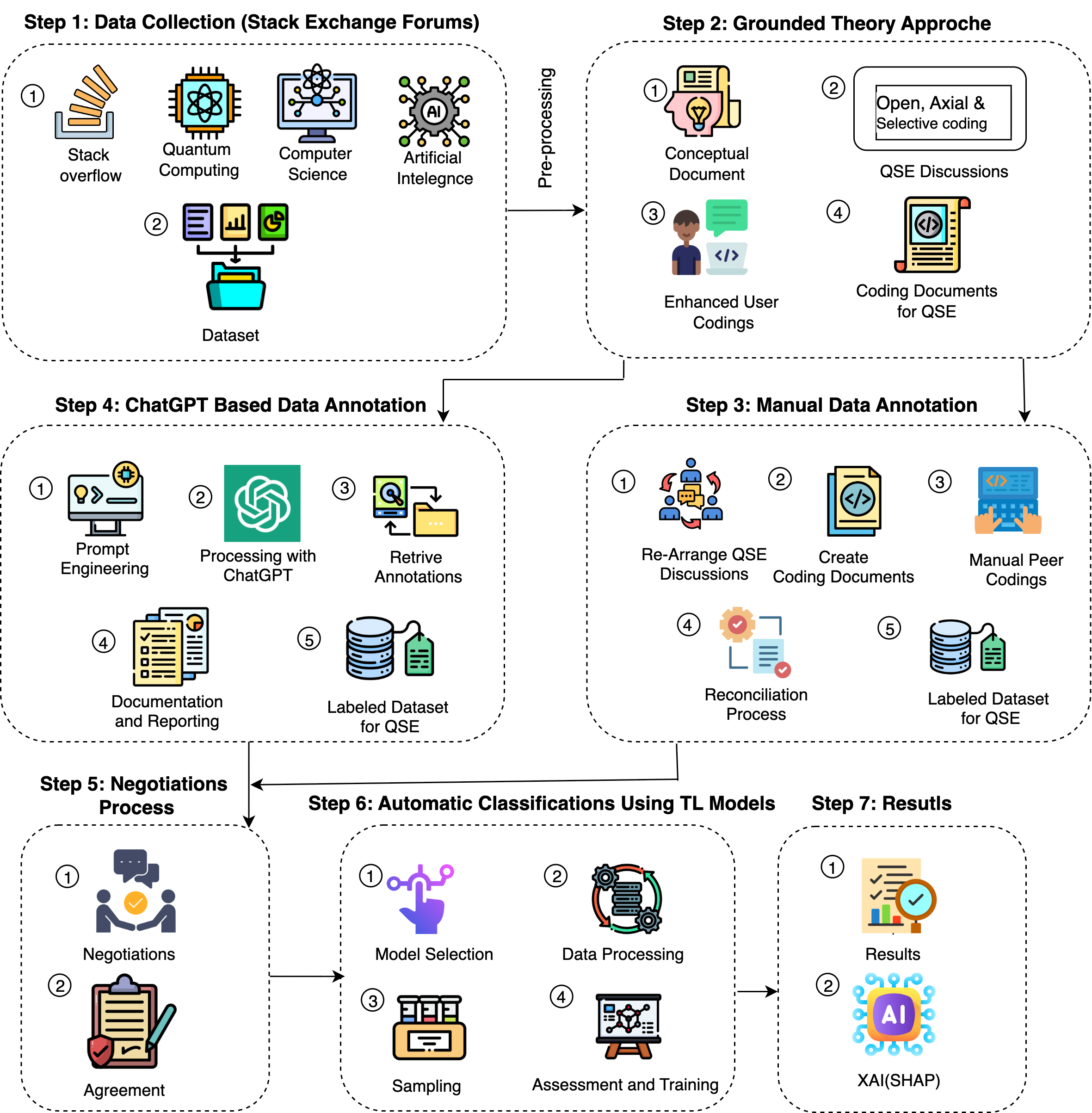}
\caption{Overview of the proposed QSE-related discussion mining approach.}
\label{fig:overview}
\end{figure*}

The proposed research methodology for classifying developers' discussions from various social media platforms into QSE challenges comprises four main phases, shown in Figure \ref{fig:overview}; each methodological step is elaborated below.

\subsubsection{Research Data Gathering and Curation}
For the proposed approach, a research dataset on QSE is developed by systematically collecting developers' discussions from SO, QCSE, CSSE, and AISE forums due to their importance in software engineering literature \cite{ahmad2018survey}, \cite{beyer2020kind}. Recently, El Aoun et al. have explored these platforms for identifying quantum-related challenges using topic modeling \cite{li2021understanding}. Also, Khan et al., \cite{khan2024insights} explored these Q\&A forums for identifying challenges encountered by software developers when implementing various software methodologies. These  Q\&A platforms prove an important source for improving various software engineering activities \cite{ahmad2018survey}. As of June 2024, SO had over 23 million, AISE had 73,124, QCSE had 23,740, and CSSE had 140,047 registered users. The SO repository, with over 24 million questions and 35 million answers, is the largest question-and-answer platform for software developers. We focused on collecting developers' discussions about quantum computing, programming languages, algorithms, and tools from these platforms, aiming to help quantum developers better identify challenges by adding a challenge type to the existing quantum tags.  

We navigated the SO, AISE, QCSE, and CSSE forums using tailored queries to extract developer discussions relevant to QSE. For SO, queries targeted questions tagged with identifiers like post-quantum-cryptography, q\#, quantum-computing, qiskit, qcl, qutip, qubit, and TensorFlow quantum, based on El-Aoun et al. approach\cite{li2021understanding}. These queries retrieved accepted and non-accepted answers, ensuring a comprehensive view of QSE discussions. Using SEDE, data from August 2008 to August 2024 were extracted. The SO had 627 questions with 282 accepted answers. For QCSE, CSSE, and AISE, queries focused on tags like programming, classical computing, and q\#. QCSE had 1806 questions with 500 accepted answers, CSSE had 379 questions with 124 accepted answers, and AISE had 17 questions with 4 accepted answers, as shown in Table \ref{tab:ml_dataset}. 

\begin{table*}[t]
\centering
\footnotesize
\renewcommand{\arraystretch}{1.2}
\caption{Dataset for TL Experiments}
\setlength{\tabcolsep}{4pt}
\begin{tabular}{p{3.5cm}p{2.5cm}p{8cm}p{2.5cm}}
\hline
\textbf{Topic} & \textbf{Total Questions} & \textbf{Tag Set} & \textbf{Data Source} \\
\hline

Stack Overflow & 
627 & 
post-quantum cryptography, q\#, quantum-computing, qiskit, qcl, qutip, qubit, TensorFlow-quantum & 
Stack Exchange \\

Quantum Computing & 
1806 & 
programming, classical computing, q\#, qiskit, cirq, ibm-q-experience, machine-learning, qutip & 
Stack Exchange \\

Computer Science & 
379 & 
quantum-computing & 
Stack Exchange \\

Artificial Intelligence & 
17 & 
quantum-computing & 
Stack Exchange \\

\textbf{Total} & 
\textbf{2829} & 
- & 
- \\
\hline
\end{tabular}
\label{tab:ml_dataset}
\vspace{-10pt}
\end{table*}

\subsubsection{Grounded theory approach }
In the previous step, developers' QSE discussions were collected from various Q\&A forums as raw data. To make this data usable, we manually selected a sample of 300 developers' discussions to critically analyze them and identify frequently occurring challenges related to QSE, and developed a grounded theory. For this purpose, we employed Corbin's grounded theory framework \cite{strauss1998basics} to develop a coding guideline, which works as an input for preparing an annotated dataset for D\&ML algorithms. Grounded theory constructs a theoretical framework based on recurring evidence from the QSE dataset. The novel ground Theory for QSE involves frequently mentioned challenges that quantum practitioners face in Q\&A forums. The frequently reported challenges identified are Tooling, Theoretical, Learning, Conceptual, Errors, and API Usage. The identified quantum concepts were verified with challenges identified by El-Aoun et al.\cite{li2021understanding} using a topic modeling approach. Compared to El-Aoun et al.\cite{li2021understanding}, we discarded three challenges: discrepancy, review, and API change, due to insufficient evidence in developers' discussions. Some concepts recur frequently; they are merged into recurring challenges. For example, discussions about unexpected results overlap with error troubleshooting and are represented with the error quantum concept. Discussions related to "API change" are grouped in the conceptual category. The "Review" challenge type can be intertwined with the SQE "Conceptual" or "API usage" categories. The quantum grounded theory process was developed iteratively with discussions by the first, second, and third authors of the paper. The document helps annotators label the raw developer discussion for D\&ML classifiers and minimize discrepancies among coders when annotating the dataset for D\&ML learning experiments.
\subsubsection{Analyzing content manually (Reviews Annotation with Humans and ChatGPT)}
After developing grounded theory, raw developer discussions from Q\&A platforms must be annotated to make a purusable dataset for D\&ML and TL classifiers. Using the content analysis approach \cite{neuendorf2017content} and a QSE grounded theory document, developers' discussions are manually annotated. For this purpose, 2829 developers' discussions on QSE collected from Q\&A platforms were added in a Microsoft Excel document to be annotated by the first and third authors of the paper using grounded theory document and content analysis approach. Moreover, content analysis is challenging and time-consuming due to human annotator involvement \cite{khan2024mining}. Therefore, to possibly automate and improve the manual annotation process, the same comments were annotated using ChatGPT to compare its performance with human annotators, as an alternative data annotation step \cite{khan2025leveraging}. Annotation forms from human coders were merged to identify interceding agreements and cohesion kappa to assess ChatGPT's applicability in annotating QSE-related comments. To resolve conflicts, another round was conducted between human coders and ChatGPT to agree on comment annotations. The detailed process is described in section 4.2.

\subsubsection{Classifying Developer Discussions using D\&ML and TL classifiers}
After annotating QSE-related developer discussions, we assessed existing fine-tuned D\&ML and TL classifiers in categorizing these discussions into QSE-related challenges identified via a grounded theory approach. These QSE challenge categories can be used as an additional tag to the existing QSE-related tags, improving QSE-related knowledge identification for software developers on the existing Q\&A platforms.  For this purpose, we performed the following: A preprocessing pipeline removed special characters, URLs, and symbols using regular expressions. NLP techniques, like stemming and removing stop words, enhanced readability for D\&ML algorithms. Feature engineering approaches, such as CountVectorizer and TFIDF, were employed to evaluate their effectiveness in classifying discussions. For deep learning classifiers, various feature values were experimented with and fine-tuned for improved results. Data-balancing methods, such as under-sampling and over-sampling, address data imbalance across various challenges. Data augmentation enhanced classifier performance, including random word deletion with a 20\% retention probability, random word swapping up to two times, and random insertion of synonyms. For the proposed approach, we removed the experimental steps for ML and DL experiments due to the length of the paper. However, the details are listed in the previously published paper \cite{husain2025exploring}. Moreover, for the transfer-learning experiments, we first selected the existing transfer-learning algorithms based on their performance on textual data analysis. Next, we perform preprocessing and fine-tuning to achieve improved results using TL algorithms. Additionally, the instances across different categories are highly imbalanced. Therefore, to balance it, we applied a data sampling approach. A cross-validation approach is used to train and validate D\&ML and transfer classifiers for generalized results. D\&ML and transformer models' performance was evaluated using the metrics of precision, recall, F1-score, ROC, and learning curves. Finally, we employed the SHAP explainable approach to explore the block-box nature of transformer algorithms and identify frequently used features by them for decision-making.

\section{PROCESSING DEVELOPERS DISCUSSION}
This section outlines the process of identifying the various challenges that quantum developers face while developing quantum-related applications. In addition, we shed light on how to annotate developer discussions collected from various Q\&A platforms into the challenges identified previously using the grounded theory approach. Below, we elaborated on the process.
\subsection{Statements of Developers}
Organizing existing developers information on these Q\&A platforms can help developers identify relevant information promptly and more accurately. QSE is a new area where quantum developers frequently face challenges \cite{li2021understanding}. Therefore, we analyze developers' discussions to identify the frequent challenges that quantum developers face, including tooling, theory, learning, concepts, errors, and API usage. These categories can be added to the existing searching algorithms for improved identification of challenge-based developer posts. The related quantum codes identified during the annotation emerged as concepts, resulting in a novel quantum theory for software developers. Concepts in the quantum coding guideline were identified by their consistent presence in discussions and relevance to the research approach. If a quantum concept appears less frequently in discussions, it merges with a related quantum concept. For example, the "API change" concept can integrate into the "Conceptual category," aligning with the definition where developers ask about the API's working. Below, we discuss the identified QSE-related concepts.

\begin{table*}[t]
\centering
\footnotesize
\caption{Quantum Concepts with Examples from Q\&A Forums}
\begin{tabular}{p{1.5cm}p{15.8cm}}
\hline
\textbf{ Concept} & \textbf{Developers' Discussion Examples} \\
\hline

Conceptual & Quantum Computing and Encryption Breaking:
"I read that Quantum Computers can break most encryption... How? I get lost at quantum bits being 1, 0, or both. Can someone explain in plain English without the math?" \\

Theoretical & NP-hard Cryptography:
"Are there public key algorithms based on NP-complete problems rather than factorization/discrete logs, in case quantum computing becomes practical?" \\

Tooling &  QASM Simulation Issues:
"Using IBM Quantum Experience: Dragging gates is slow, no Toffoli gate. Need QASM editor/documentation. Found these links but unclear which QASM version IBM uses..." \\

Learning & Book Request:
"Looking for a book covering Shor's algorithm, McEliece cryptosystem, Lattice-based crypto, and Discrete logarithms. Any recommendations?" \\

Error & Q\# Build Problem:
"Getting error QS1001: Microsoft.Quantum.Canon.dll not found. Missing .nuspec file. Nuget restore didn't help. Solutions?" \\

API Usage & Break Equivalent in Q\#:
"How to exit a loop when condition met? In C\#: if(i==3) break. What's the Q\# equivalent?" \\
\hline
\end{tabular}
\label{tab:quantum_concepts}
\end{table*}

\subsubsection{Conceptual}
The code "Conceptual" is assigned to developers' discussions on Q\&A platforms focusing on understanding the background, limitations, and concepts of quantum programming APIs. Analysis shows that quantum developers discuss quantum computing history or debate the feasibility of certain concepts. Researchers categorize discussions into conceptual categories if questions contain terms like why, is possible \cite{beyer2014manual}, what \cite{rosen2016mobile}, how/why \cite{allamanis2013and}, and understanding concepts \cite{beyer2020kind}. The definition for categorizing discussions aligns with approaches by Beyer and Pinzger \cite{beyer2014manual}, Rosen and Shihab \cite{rosen2016mobile}, Allamanis and Sutton \cite{allamanis2013and}, and Beyer et al. \cite{beyer2020kind}. An example is in Row 2 of Table \ref{tab:quantum_concepts}.

\subsubsection{Theoretical} The code “Theoretical” is assigned to Q\&A platform discussions on quantum programs, algorithms, and principles. Developers may inquire about the mathematical foundations of quantum mechanics or seek explanations for quantum algorithms. For instance, a developer might ask how superposition or entanglement affects quantum algorithm design, as shown in Table \ref{tab:quantum_concepts}, row 3. It discusses the theoretical implications of practical quantum computing on public-key cryptographic algorithms and complexity theory. El-Aoun et al.\cite{li2021understanding} identified that developers often discuss theoretical problems related to quantum computing in Q\&A discussions using a topic modelling approach.
\subsubsection{Tooling} This category pertains to developer discussions on tools and software usage in quantum development. The code "tooling" is assigned to developers' discussions on Q\&A platforms that discuss possible quantum tools, seek expertise, or help with quantum-related tools. Quantum computing is an emerging discipline where vendors develop new tools to solve quantum-related problems. During annotation, quantum developers often seek guidance on using specific software tools or frameworks for quantum development. An example is shown in Table \ref{tab:quantum_concepts}, Row 4. This category aligns with the El-Aoun et al. \cite{li2021understanding} approach, which identified "Tooling" as a frequently occurring quantum challenge developers discuss in Q\&A forums.
\subsubsection{Learning} In the developer discussions dataset where quantum practitioners seek resources about quantum computing from the Q\&A community, we assigned the code “learning.” This category includes requests for learning resources, tutorials, and references for quantum computing. Developers often seek recommendations for books, online courses, or other resources to enhance their understanding of quantum programming languages and algorithms. For instance, a quantum developer inquires about the best resources for learning Q\# or mastering Grover's search algorithm, as shown in Row 5 of Table \ref{tab:quantum_concepts}. The “Learning” category for the quantum classification taxonomy aligns with Beyer et al. \cite{beyer2020kind} and El-Aoun et al. \cite{li2021understanding} “Learning” category identified for android-related topics and quantum challenges using topic modelling, respectively. Also, Allamanis and Sutton \cite{allamanis2013and} identified learning a language/Technology category with different purposes and intentions.

\subsubsection{Errors} Quantum computing is a new research and development area with limited expertise in quantum programming compared to traditional paradigms. We identified that quantum developers often seek help resolving errors while developing quantum applications. We assigned the code "errors" to developers' posts on Q\&A platforms where quantum practitioners ask for solutions to errors. During manual annotation, we found that quantum developers often encounter challenges related to quantum gates, qubit decoherence, and compatibility issues with quantum hardware. For example, a practitioner asked for help debugging quantum circuits or understanding error messages from a quantum simulator, as shown in row 6 of Table \ref{tab:quantum_concepts}. This category aligns with Beyer et al.\cite{beyer2020kind}, El-Aoun et al. \cite{li2021understanding} "Errors", and Beyer and Pinzger \cite{beyer2014manual} "Error and Exception Handling" category.

\subsubsection{API Usage} As quantum computing evolves, numerous APIs have been developed for various functionalities that quantum developers can utilize. The code “API usage” is assigned to developers' discussions in the Q\&A seeking help in efficiently implementing a particular Quantum-related API. During manual annotation, it was noted that discussions focused on using APIs for quantum programming tasks. Developers seek guidance on integrating quantum computing platforms and services into their projects using APIs. For example, a developer may inquire about best practices for interfacing with the IBM Quantum Experience API or accessing the Rigetti Forest API for quantum program execution, as shown in Table \ref{tab:quantum_concepts}, row 7. This category aligns with those of El-Aoun et al.\cite{li2021understanding} and Beyer et al.\cite{beyer2020kind} API usage category, Allamanis and Sutton’s \cite{allamanis2013and} “Do not work” category, and Beyer and Pinzger's \cite{beyer2014manual} “How to” category.

\subsection{Developers Statement Labeling}
To annotate developer discussions from Q\&A platforms, we used content analysis \cite{neuendorf2017content} and QSE  grounded document, examining each developer's question to identify their quantum challenge category. The goal is to develop a truth set for training and testing the fine-tuned D\&ML and transfer algorithms to categorize developer questions into quantum categories automatically. The annotation process is described below:

\subsubsection{Annotation of Developer Discussions}
\begin{figure}[h]
  \centering
  \includegraphics[width=\linewidth]{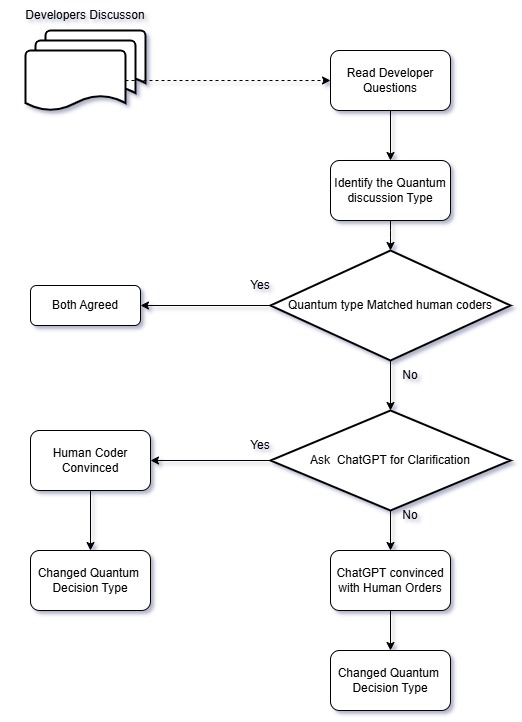}
\caption{Annotation and negotiation process with ChatGPT.}
\label{fig:annotations}
\end{figure}

Annotation was conducted in two steps. First, the first and third authors individually annotated 2829 developers' discussions from Q\&A forums. Second, ChatGPT annotated the same dataset to compare its performance with human coders. Human annotators were given a coding guideline and form with developer comments. Coders assessed the title and content to determine the quantum category: tooling, theoretical, learning, conceptual, errors, or API Usage. Figure \ref{fig:annotations} shows an overview of the coding process. Developers' questions were grouped by platform, including answers for clarity. The annotation process was iterative, as shown in Figure 1. The average annotation time was 44 working hours. The coders start and resume annotation at any time. After individual annotation, the results were combined to analyze intercoder disagreements. The intercoder agreement was 90.81\%, with Cohen's kappa at 0.46 (moderate agreement), with 95\% confidence intervals [0.43, 0.49]. Conflicts were resolved through discussion, with a senior software researcher (second author) intervening for persistent disagreements.
To validate the annotation process, we used ChatGPT to annotate developers' discussions into quantum categories and compared its performance with human annotators. We trained ChatGPT with definitions of quantum challenge types from novel quantum-grounded theory using the existing prompt. We supplied each developer's discussion to ChatGPT and identified its possible quantum challenge type with a rationale. We recorded the number of discussions where manual annotators and ChatGPT agreed and disagreed. Table 5 shows detailed annotation by human coders and ChatGPT, agreeing on 2547 discussions for the same quantum challenge category and disagreeing on 282, resulting in potential conflicts. The Cohen's kappa was 0.46, indicating moderate agreement between human coders and ChatGPT on the Cohen's kappa scale.

\begin{table*}[t]
\centering
\footnotesize
\caption{Summary of Annotated Developer Discussions on Quantum Computing}
\label{tab:quantum_discussions}
\begin{tabular}{p{2.5cm}p{12.2cm}p{1.3cm}}
\hline
\textbf{Q\_Title} & \textbf{Q\_description} & \textbf{Q\_Type} \\ 
\hline

Quantum computing \& encryption break & 
I read that Quantum Computers can break most encryption in minutes. How? I get lost at quantum bits being 1, 0, or both. Can someone explain in plain English without the math? &
Conceptual \\

Qubit density to Bloch vector & 
Given a 2x2 density matrix, how to compute the Bloch sphere point? State |0⟩-|1⟩ has matrix [[0.5,-0.5],[-0.5,0.5]] and should be on X axis. Matrix [[0.5,0],[0,0.5]] should be at origin. &
Theoretical \\

What is Quantum Computing? & 
In physics, particles in multiple states simultaneously. In computing, bits as 1, 0, both, or NULL? Applications to processors, programming, security? Any practical implementations? &
Learning \\

floor/ceil in QCL & 
What do QCL's floor() and ceil() operators do exactly? They seem related to math operations but need clarification. &
Tooling \\

QuTiP integration failure & 
Getting errors solving Lindblad equation. Increasing nsteps doesn't help. Changing time values (np.linspace) has no effect. Need solution. &
Errors \\

Can Forest crack crypto? & 
Can Rigetti Forest break public-key crypto (e.g., Bitcoin) in reasonable time? If yes, show solution using pyQuil. &
API Usage \\
\hline
\end{tabular}
\end{table*}

\begin{table}[t]
\centering
\footnotesize
\renewcommand{\arraystretch}{1.2}
\caption{Details of Human and ChatGPT Annotations}
\label{tab:annotations}
\begin{tabular}{p{3cm}p{3cm}p{2cm}}
\hline
\textbf{Quantum Category} & \textbf{Human Freq} & \textbf{ChatGPT Freq} \\
\hline
Tooling & 600 & 611 \\
Theoretical & 414 & 446 \\
Learning & 159 & 158 \\
Conceptual & 606 & 597 \\
Errors & 803 & 809 \\
API Usage & 247 & 208 \\
\hline
\multicolumn{3}{l}{\textbf{Total Number of Questions:} 2829 (Human), 2829 (ChatGPT)} \\
\hline
\multicolumn{3}{p{8cm}}{
\textbf{Disagreement Analysis:} There are 282 disagreements between human coders and ChatGPT. The number of learning experiences aligns with this count. In each category, there are disagreements where the human coders, for example, incorporate the learning category where ChatGPT utilizes tools. Overall, disagreements exist across various categories.} \\
\hline
\end{tabular}
\end{table}

A negotiation process resolved conflicts between human coders and ChatGPT. For conflicting developer discussions on the Q\&A forum, a second round with ChatGPT was initiated. A generic query asked for elaboration on selected quantum categories. This discussion continued until a consensus was reached. Among 282 conflicting discussions, ChatGPT agreed with human annotators on 173 questions after reasoning. Conversely, human coders agreed with ChatGPT on 109 developer feedback after receiving justification. This process curated a conflict-free dataset for QSE, as shown in Table \ref{tab:quantum_discussions}. This step enhances the labelled dataset's quality by validating human annotation with ChatGPT, improving the effectiveness of subsequent D\&ML and TL classifiers in identifying quantum developer challenge types.

\begin{table*}[t]
\begin{promptbox}[Validation and Reconciliation of Annotations]\label{box:box2}
\textbf{Overview of Disagreements:} Approximately 10\% of developer discussions (282 out of 2,829) showed disagreements between human and ChatGPT annotations, mainly in boundary cases between \textit{Tooling}, \textit{Learning}, and \textit{Conceptual} categories.

\textbf{Resolution through Negotiation:} In 38.7\% of cases (109), human coders revised their labels to match ChatGPT’s reasoning; in 61.3\% (173), ChatGPT adjusted its label after reviewing human justification.

\textbf{Example of Review Disagreement and Resolution:}
\begin{lstlisting}
- User Review: "It's been a while since I got into QCL but I ran into some difficulties by trying to figure out the functioning of measure regX or measure regX,m. What does this do on the quantum registers? And the m integer?"
- ChatGPT Annotation: Learning (intent to understand functionality).
- Human Annotation: Tooling (focus on QCL operator).
- Outcome: Humans agreed with ChatGPT after recognizing the query's educational intent. Final label: Learning.
\end{lstlisting}

\textbf{Implication:} This human-AI collaboration improves annotation quality by combining expert judgment with scalable pattern recognition, resulting in a more accurate and reliable dataset for training classifiers in quantum software engineering.
\end{promptbox}
\end{table*}

\begin{table*}[t]
\begin{promptbox}[Prompts for QSE Challenges Detection Using ChatGPT]\label{box:prompt_example}
\textbf{Example of QSE Categorization Prompt:}
\begin{lstlisting}
"Please categorize the following quantum developer discussion based on these challenge categories:
- Conceptual: Understanding the basic concepts of quantum programming and computing.
- Theoretical: Inquiries regarding the mathematical and theoretical foundations of quantum computing.
- Tooling: Issues related to the software tools, APIs, and quantum simulators used in quantum development.
- Learning: Requests for learning resources, tutorials, or guidance on quantum algorithms and programming languages.
- Errors: Problems encountered during the development process, such as errors in code or simulation issues.
- API Usage: Assistance with integrating or using quantum-related APIs in development.
\end{lstlisting}
\end{promptbox}
\end{table*}
\subsection{Frequency of the Developer’s Discussions on Q\&A Platforms}
Before subjecting the developer discussions to fine-tuned transformer algorithms, it is essential to identify the frequency of developer instances across each quantum challenge category to achieve better results. A detailed analysis aids in understanding the data, quantum trends, challenges, and recent developments. We utilized content analysis to examine the frequencies of developer discussions across categories, providing insights for quantum software vendors and developers. The goal is to introduce various types of quantum discussions, including tooling, theoretical, learning, conceptual, error analysis, and API usage, to assist quantum vendors in understanding trends and difficulties faced by practitioners. The most frequent type of developer question on Q\&A platforms is the "Error" quantum type, comprising 28.80\% of discussions (815 questions), as shown in Table 6. Quantum software development is emerging, with limited resources and tools. Developers often encounter challenges and seek assistance from open-source communities, primarily related to troubleshooting, debugging, and addressing technical issues in quantum software. The second most frequently reported quantum challenge is conceptual, accounting for 21.56\% of discussions (610 instances). Quantum researchers, investors, and vendors are increasingly interested in proposing new research methodologies and techniques for quantum problems. Developers engage in conceptual discussions to understand the theoretical and fundamental principles behind software applications. Practitioners discuss various concepts related to quantum programming, debating the feasibility of certain quantum concepts. This presents opportunities for software vendors and investors to consider practitioners' discussions about the effectiveness of proposed quantum methodologies and overcome challenges for better productivity. The third most important category identified through content analysis is Tooling, constituting 21.06\% (596) of the total discussions (2829 developers' questions). Quantum developers frequently seek information on tools and software usage in quantum programming. As researchers introduce new tools to automate quantum functionalities, practitioners are likely to turn to open-source developers to share their experiences. This suggests that developers are actively exploring tools to enhance quantum software development. It presents an opportunity for quantum vendors to concentrate on tool usage and assist developers in better understanding the necessary functionalities. Such feedback could enhance the usability and developer experience with the available quantum tools. The next significant category is theoretical, accounting for 14.66\% (415 developers' threads) of the total discussions (2829), as shown in Table \ref{tab:final_dataset}. Quantum developers engage in Q\&A platforms to discuss complex theoretical concepts, algorithms, and principles for quantum-based applications. Quantum vendors can incorporate essential information and examples to improve understanding. The second least mentioned quantum challenge type is API Usage, making up 8.02\% (227 developers' discussion questions). Quantum software development is intricate. Creating such applications from scratch is time-consuming and resource-intensive. Quantum developers often seek guidance on integrating APIs, reporting challenges in understanding documentation, and troubleshooting API-related issues. This helps quantum investors and vendors recognize the potential value in developing free APIs that extend quantum functionalities, saving developers' time and resources. Finally, learning emerged as the least-discussed quantum challenge type, comprising only 5.86\% (166) of the total discussions (2829 questions). Quantum developers seek recommendations for resources to enhance their understanding of quantum programming languages, algorithms, and related concepts. Identifying such information can help software quantum vendors provide assistance on frequently occurring errors by embedding it into the existing quantum tools. Similarly, vendors can support the often-asked software quantum concepts in developer platforms. Moreover, crash courses can be offered on emerging technologies, such as software quantum computing, to boost software productivity.

\begin{table}[t]
\centering
\footnotesize
\renewcommand{\arraystretch}{1.2}
\caption{Finalized Conflict-Free Annotated Quantum Dataset}
\label{tab:final_dataset}
\begin{tabular}{lc}
\hline
\textbf{Quantum Category} & \textbf{Number of Developers' Discussions} \\
\hline
Tooling & 596 \\

Theoretical & 415 \\

Learning & 166 \\

Conceptual & 610 \\

Errors & 815 \\

API Usage & 227 \\
\hline
\textbf{Total} & \textbf{2829} \\
\hline
\end{tabular}
\end{table}

\section{Automated classifications of developer discussions Using TL models}
In the literature, mining developers' feedback on Q\&A platforms significantly contributes to software evolution\cite{ahmad2018survey}. Researchers have mined Q\&A platforms, particularly SO, for various purposes, such as developers' opinions about APIs \cite{uddin2019automatic}, mining quantum programming \cite{khan2025mining}, identifying challenges for software methodologies and popular topics \cite{khan2024insights}, popular programming language developers \cite{baquero2017predicting}, and bug severity prediction \cite{tan2020bug}. Manually identifying useful information for software developers and vendors is cumbersome and resource-intensive \cite{ahmad2018survey}. Complementary to these approaches, we mined developer discussions from Q\&A platforms to identify frequently mentioned quantum challenges, aiming to provide opportunities for quantum vendors to focus on key areas developers struggle with or discuss most frequently. Also, the proposed approach can improve the searching mechanism of existing Q\&A platforms by using the proposed classification approach to identify the category type and add to existing tags. To achieve a more robust and context-aware classification, in addition to the existing D\&ML experiments, we experimented with the state-of-the-art TL algorithms to improve accuracy and low ratings of false positives and negatives. These advanced models are better suited for understanding the complex linguistic nuances present in the developer discussions. Moreover, we do not include the preprocessing, fine-tuning, data balancing, and detailed explanation of D\&ML experiments results, as it is listed in the previously published paper. Therefore, in this paper, we focused more on the transformer-based experiments. Below, we elaborate on the experimental setup for applying these TL models and their subsequent steps.

\subsection{Experimental setup}
To ensure the reliability and reproducibility of the results, a robust computational environment was configured for all data processing and model training tasks. The primary development was conducted on a Macbook Pro M2, with experiments run within a Jupyter notebook environment. This setup allowed for iterative development and seamless integration with specialized Python libraries for NLP.

The software stack leveraged state-of-the-art libraries such as PyTorch was used as primary framework for building and trainings TL architectures, while the Hugging Face Transformers library provided the core implementation for the models and their tokenizers. For evaluation Scikit-learn was employed to calculate performance metrics and implement cross-validation splits. All data manipulations and preparation was handled using Pandas and NumPy, and visualizations were generated with Matplotlib and Seaborn. This setup ensured optimal and consistent conditions for the computationally intensive task required by this study. A complete summary of system configurations is presented in Table \ref{tab:system_config}

\begin{table}[t]
\centering
\caption{System Configuration}
\label{tab:system_config}
\begin{tabular}{l l}
\toprule
\textbf{Specification} & \textbf{Details} \\
\midrule
Hardware Accelerator & NVIDIA CUDA-enabled GPU \\
Development Environment & Jupyter Notebook \\
Programming Language & Python \\
Machine Learning Framework & PyTorch \\
NLP Library & HuggingFace Transformers \\
Data Manipulation & Pandas, NumPy \\
Visualization & Matplotlib, Seaborn \\
Evaluation Tools & Scikit-learn \\
\bottomrule
\end{tabular}%
\end{table}

%For the D\&ML experiments, we used text preprocessing and feature engineering to evaluate classifiers and their effectiveness in identifying developer discussions in Stack Exchange forums. We selected D\&ML algorithms proven effective with textual data \cite{khan2022valuating} \cite{khan2024mining}, specifically RNN, GRU, LSTM, CNN, FNN, and MLP. These algorithms were used to classify developer questions from Q\&A forums and compare their results on classifying discussions into various quantum types. Experiments were conducted in Python, utilizing libraries and tools suited for D\&ML tasks to ensure robust results.

\subsection{TL Models Selection}
Selecting the appropriate classification model is crucial for improved and generalized results. This study employs three state-of-the-art TL models based on the Transfer Architecture: BERT, DistilBERT, and RoBERTa. These models were chosen for their robust performance on a wide range of NLP benchmarks. Their pretrained nature allows them to leverage deep linguistic knowledge to effectively discern the context and semantics within the specialized domain of QSE forums.
\subsubsection{BERT} 
As a foundational model, BERT \cite{devlin2019bert} TL algorithm is selected to serve as a baseline, which proves to be a better choice compared to existing ML approaches \cite{gonzalez2020comparing}. Its key innovation lies in its bidirectional training, which allows it to process the entire context of a text simultaneously, rather than sequentially. This enables a deeper understanding of linguistic relationships within a sentence. This makes it a better choice for processing SQE-related developers' discussion posts, which contain lengthy textual information at times. For this study, the Bert base-uncased configuration was utilized, which consists of 12 encoder layers and a 768-dimensional hidden state. Its pre-training on large, general-purpose corpora provides a solid foundation for fine-tuning on the specialized vocabulary found in QSE discussions.
\subsubsection{DistilBERT}
DistilBERT was included to evaluate the performance of a more computationally efficient architecture \cite{sanh2019distilbert}. This model is a condensed version of BERT, created using a technique known as knowledge distillation. It is approximately 40\% smaller and significantly faster during inference while retaining the vast majority of BERT's performance. This makes it a compelling candidate for real-world engineering applications where resource constraints and response time are important considerations. Its architecture, while more compact, maintains the core principles of the Transformer, allowing for a meaningful comparison of the performance-efficiency trade-off.
\subsubsection{RoBERTa}
To evaluate a model featuring an improved pre-training methodology, RoBERTa was selected for this study. Developed by Facebook AI RoBERTa modifies the original BERT pre-training process by removing the next-sentence prediction objective and instead training exclusively with a masked language modeling (MLM) objective on a significantly larger text corpus \cite{r18}. Furthermore, it employs a dynamic masking strategy, where the masking patterns applied to the input tokens are varied during each training epoch. For tokenization, RoBERTa utilizes a Byte-Pair Encoding (BPE) vocabulary, which is particularly effective at handling the specialized technical terms found in QSE developer forums. These optimizations result in a model with a more nuanced understanding of language, making it exceptionally well-suited for fine-grained classification tasks of distinguishing between the subtle differences in developer discussions.
\subsection{Data Preparation and Pre-processing for TL models}
The preparation of the raw textual data from developer forums is crucial for effective analysis with TL models. This stage involved a systematic workflow to clean, format, and structure the data to align with the input requirement of the transfer architectures, ensuring that the semantic integrity of the developer discussions was preserved while optimizing them for the classification task.
\subsubsection{Class Label Conversion}
For the supervised classification task, the six categorical labels representing the QSE challenges (\textit{Tooling, Conceptual, Errors, Theoretical, API Usage,} and \textit{Learning}) were converted into numerical format. This is achieved by mapping each unique category string to a distinct integer index (0 through 5). This transformation is a standers requirement for DL and TL frameworks like PyTorch, as it allows the model's loss function to compute gradients effectively during training. 
\subsubsection{Model-Specific Tokenization}
Unlike traditional NLP that requires aggressive cleaning, Transformer models utilize their own specialized tokenizers that are trained to handle near-raw text. The primary Pre-processing step, therefore, was a basic cleaning pass to remove artifacts like HTML tags or stray URLs. Subsequently, the cleaned text of each developer discussion was processed using the specific tokenizer corresponding to each model (i.e., \textit{BertTokenizerFast}, \textit{RobertaTokenizerFast}, and \textit{DistilBertTokenizerFast}).
 This process involves three key actions: (1) converting the text into subword tokens based on the model vocabulary,(2) adding special tokens (e.g., \textit{CLS} and \textit{SEP}) required by the architecture, and (3) padding or truncating all sequences to a uniform length to enable batch processing. Attention masks were also generated to allow the model to differentiate between meaningful tokens and padding tokens.
 \subsubsection{Token Length Distribution Analysis}
\begin{figure}[h]
  \centering
  \includegraphics[width=\linewidth]{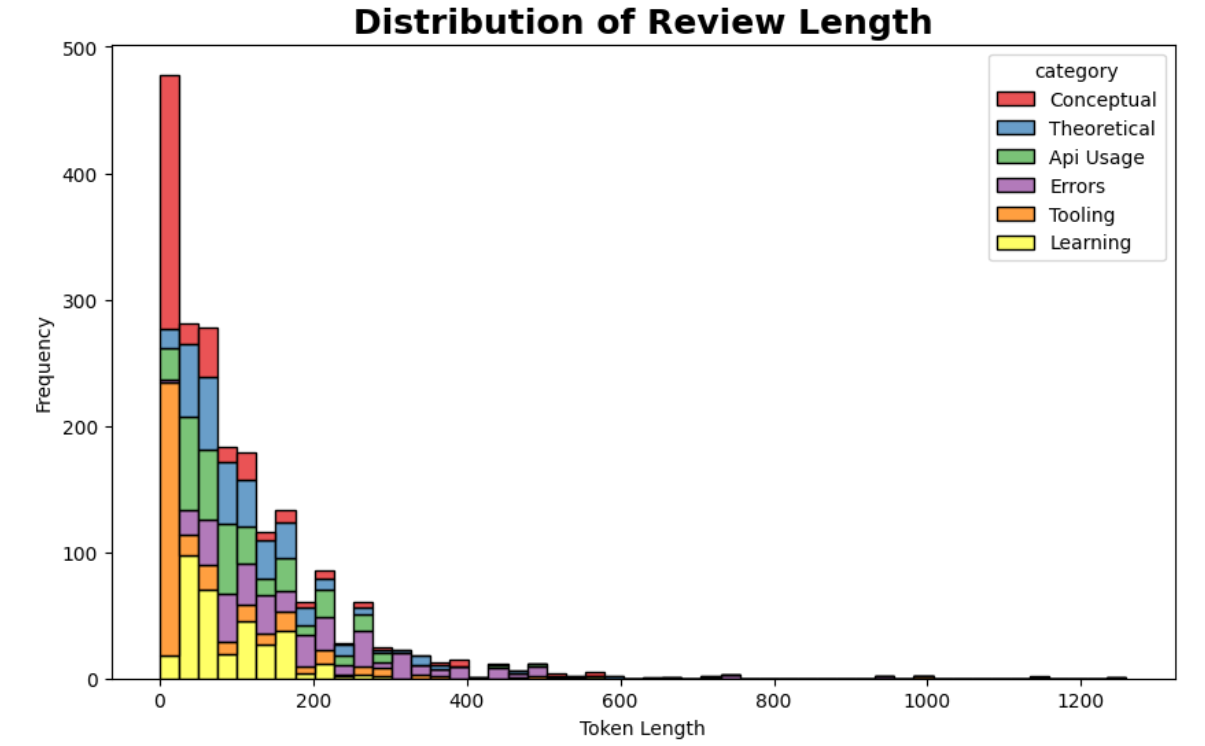}
\caption{Token length distribution across all categories.}
\label{fig:token}
\end{figure}

 To determine an optimal sequence length for padding and function, an initial analysis of the token length across all developers' discussions was conducted. The distribution revealed that the vast majority of the content could be effectively captured within a sequence length of 128 tokens. This value was chosen as the max\_length parameter for tokenization, as it provides an excellent balance between retaining crucial textual information and maintaining computational efficiency during the model training and inference phases. The histogram in the Figure \ref{fig:token} highlights the right-skewed nature of the distribution, with most discussions falling below 200 tokens but a significant tail extending beyond that. 
\subsection{Data Sampling}
\begin{figure}[h]
  \centering
  \includegraphics[width=\linewidth]{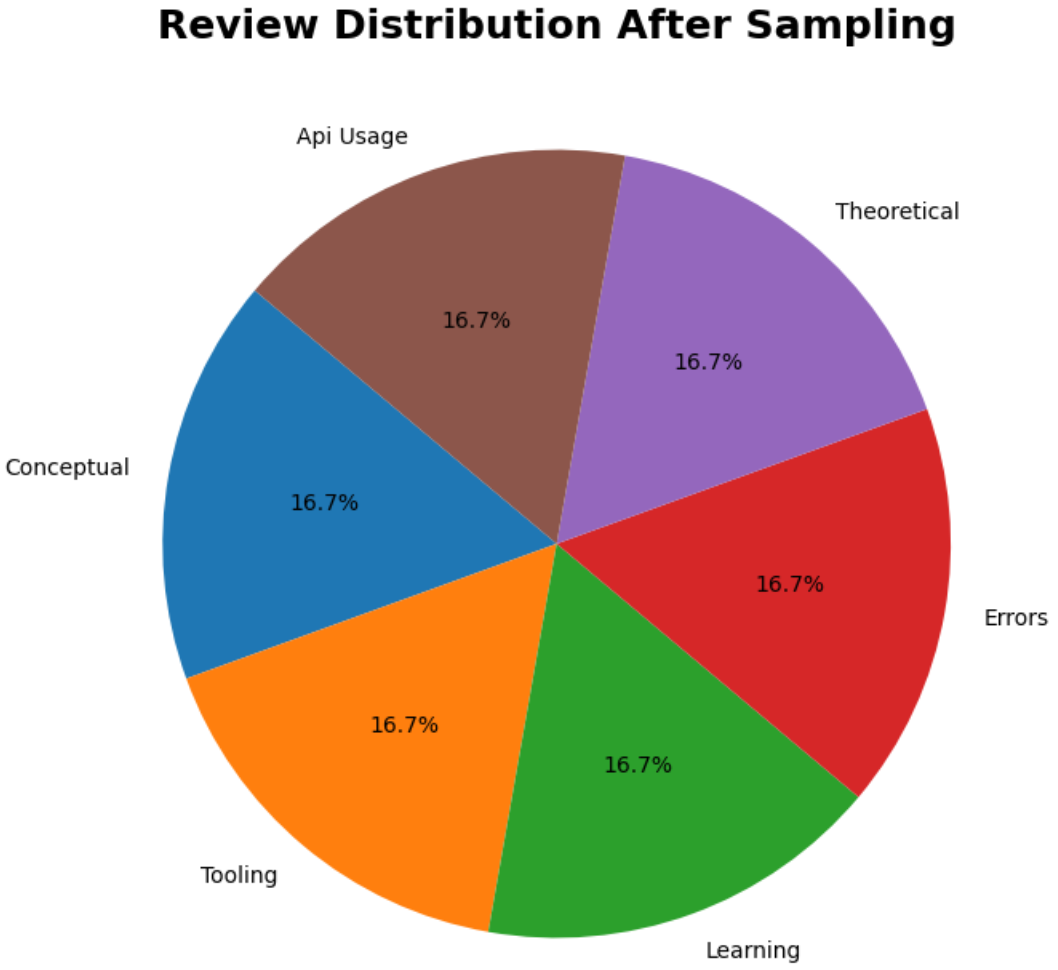}
\caption{Categories distribution after data balancing.}
\label{fig:balancing}
\end{figure}

The initial conflict-free dataset, while rigorously validated, exhibited a significant class imbalance, which is a common challenge in supervised ML \cite{khan2024mining, khan2025leveraging}. An analysis of the class distribution, as shown in the Table \ref{tab:final_dataset}, revealed that some QSE challenge categories were far more prevalent than others. For instance, the 'Errors' category contained 815 developer discussions, whereas the 'Learning' category comprised only 166. Training a model on such an imbalanced dataset can introduce significant bias, causing the classifier to favor the majority classes while performing poorly on the underrepresented minority classes.
\\
To mitigate this potential bias and ensure the model learns the features of each category with equal importance, a strategic up-sampling approach was employed. This technique balances the dataset by randomly duplicating instances from the minority classes until they match the sample count of the majority class. The \textit{scikit-learn} library \textit{resample} function was used to perform this task. The result of this procedure is a perfectly class-balanced training dataset, as illustrated in the distribution shown in Figure \ref{fig:balancing}. This balanced dataset forms the foundation for training the TL models, enabling them to develop a more robust and generalizable understanding of all six QSE challenge categories.
\subsection{Training and Fine-tuning of TL models}
\begin{figure*}[h]
  \centering
  \includegraphics[width=\linewidth]{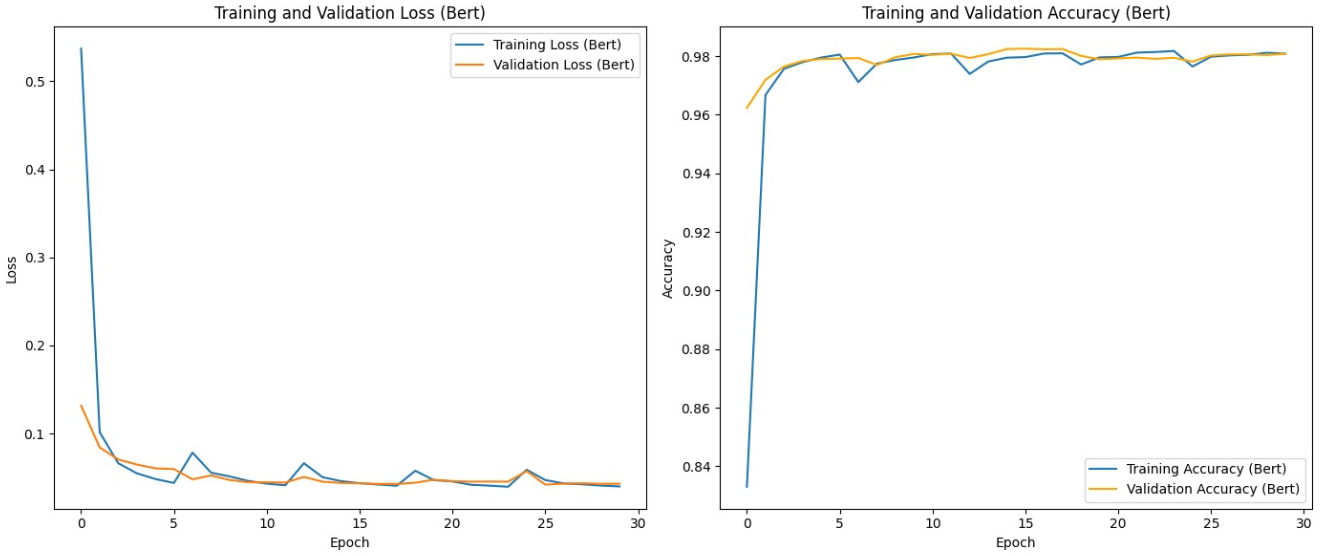}
\caption{Training and validation accuracy and loss for BERT model.}
  \label{fig:BERT}
\end{figure*}

The pre-trained models were not used off-the-shelf: they were fine-tuned on the domain-specific QSE challenge dataset to adapt their general linguistics knowledge to the specific vocabulary and context of developer forums. To ensure the results are robust and to prevent overfitting, a 5-fold cross-validation strategy was employed. For each fold, the dataset was split into a training set (80\%) and a validation set (20\%), ensuring that every data point was used for both training and validation across the five iterations.
\\
The fine-tuning process of each model was conducted over 30 epochs with a batch size of 16. We utilized the AdamW optimizer with a learning rate of 2e-5, which is a standard and effective choice for training Transformer models. A linear learning rate scheduler with a warmup phase was also implemented to help stabilize the training process in the initial epochs.
\\
The training and validation progress for each model was carefully monitored. The learning curves for the BERT model, presented in Figure \ref{fig:BERT}, show a consistent decrease in loss and an increase in accuracy, indicating stable learning. In addition, all three models were successfully fine-tuned on the dataset without significant signs of overfitting.
\subsection{Assessment and Training}
To train the TL models and evaluate their performance, a 5-fold cross-validation strategy was implemented. In this approach, the dataset is partitioned into five equal-sized folds. During each iteration, one fold is held out for validation and the remaining four folds are used for training. This process is repeated five times , with each fold serving as the validation set exactly once. This method ensures that every data points is used for both training and validation , providing a more reliable estimate of the model generalization performance that a single train-test split.
\\
The effectiveness of the classifiers was assessed using a standards set of performance metrics:Precision, Recall, and F1-Score. These matrices were calculated for each of the QSE challenge categories.
Precision ($P_k$) measures the accuracy of the positive predictions, indicating what proportion of instances classified as a certain category truly belong to that category. It is calculated as:
\begin{equation}
P_k = \frac{TP_k}{TP_k + FP_k}
\end{equation}
where $TP_k$ are the true positives and $FP_k$ are the false positives for class $k$.
\\
Recall ($R_k$), also known as sensitivity, measures the model's ability to identify all relevant instances of a class. It is calculated as:
\begin{equation}
 R_k = \frac{TP_k}{TP_k + FN_k}
\end{equation}
where $FN_k$ are the false negatives for class $k$.
\\
The F1-Score ($F1_k$) provides a single, balanced measure of a model's performance by calculating the harmonic mean of Precision and Recall. This is particularly useful for imbalanced datasets. The formula is:
\begin{equation}
F1_k = 2 \times \frac{P_k \times R_k}{P_k + R_k}
\end{equation}

  \section{Results}
 This section presents the performance evaluation of TL-based and D\&ML classifiers in categorizing QSE challenges using various experimental settings. As shown in Table~\ref{tab:classifier_performance_updated}, the TL models BERT and DistilBERT achieved an overall accuracy of 95\% without any data augmentation, demonstrating their strong generalization capability in low-resource, domain-specific settings. Both models exhibited exceptional performance across all categories. BERT achieved a perfect F1-score of 0.99 in \textit{Theoretical} and \textit{API Usage}, while DistilBERT matched this level of performance in \textit{Learning}, \textit{Tooling}, and \textit{API Usage}. These results confirm that pre-trained language models can effectively capture the semantic nuances of developer discussions in QSE without requiring synthetic data enhancement.
 
Compared to fine-tuned TL algorithms, the performance of DL models was highly dependent on data augmentation, as it can be seen in Table~\ref {tab:classifier_performance_updated} that the performance of the fine-tuned CNN classifier is comparatively lower when exposed to actual developers' discussions on Q\&A forums.  The CNN model showed a significant improvement when trained with augmented data, achieving an overall accuracy of 86\%, compared to only 51\% without augmentation. This trend is consistent across categories: in \textit{Learning}, CNN’s F1-score increased from 0.30 to 0.88; in \textit{Errors}, it rose from 0.77 to 0.94; and in \textit{API Usage}, it jumped from 0.20 to 0.86. Similarly, the fine-tuned MLP model works well with augmentation, increasing its overall accuracy from 54\% to 67\%. Notably, in \textit{Errors}, MLP improved its F1-score from 0.78 to 0.84, and in \textit{Tooling}, from 0.47 to 0.65. We demonstrated the performances of CNN and MLP classifiers, while the detailed results of several D\&ML algorithms are elaborated in our previously published paper \cite{husain2025exploring}.

These findings highlight a fundamental difference in model behavior: TL models like BERT and DistilBERT are effective out-of-the-box due to their extensive pre-training on large corpora, making them robust even with limited or imbalanced datasets. On the other hand, traditional DL models such as CNN and MLP rely heavily on data augmentation techniques, including random word swapping, deletion, and synonym insertion, to learn meaningful representations from sparse real-world forum data. Overall, while both TL and augmented DL approaches can achieve high performance, the superior results of TL models without augmentation make them a more practical, scalable, and efficient solution for the automated analysis of developer discussions in emerging domains, such as QSE.

\begin{table*}[htbp]
\centering
\footnotesize
\renewcommand{\arraystretch}{0.92}
\setlength{\tabcolsep}{3.5pt}
\caption{Performance of TL and DL Classifiers in QSE (With and Without Data Augmentation)}
\label{tab:classifier_performance_updated}
\begin{tabular}{p{3.8cm} p{3.7cm} p{2.1cm} p{2.1cm} p{2.1cm} p{3.4cm}}
\hline
\textbf{QSE Concept} & \textbf{Classifier} & \textbf{Precision} & \textbf{Recall} & \textbf{F1-score} & \textbf{Data Augmentation} \\
\midrule
\multirow{6}{*}{Conceptual} 
    & BERT         & 0.99 & 0.94 & 0.97 & No \\
    & DistilBERT   & 0.80 & 0.99 & 0.89 & No \\
    & CNN          & 0.91 & 0.45 & 0.61 & Yes \\
    & CNN          & 0.37 & 0.36 & 0.36 & No \\
    & MLP          & 0.62 & 0.56 & 0.57 & Yes \\
    & MLP          & 0.39 & 0.54 & 0.45 & No \\
\midrule
\multirow{6}{*}{Theoretical} 
    & BERT         & 0.99 & 0.99 & 0.99 & No \\
    & DistilBERT   & 0.99 & 0.78 & 0.88 & No \\
    & CNN          & 0.76 & 0.78 & 0.74 & Yes \\
    & CNN          & 0.35 & 0.40 & 0.38 & No \\
    & MLP          & 0.56 & 0.57 & 0.54 & Yes \\
    & MLP          & 0.42 & 0.23 & 0.30 & No \\
\midrule
\multirow{6}{*}{Learning} 
    & BERT         & 0.89 & 0.99 & 0.89 & No \\
    & DistilBERT   & 0.99 & 0.99 & 0.99 & No \\
    & CNN          & 0.90 & 0.87 & 0.88 & Yes \\
    & CNN          & 0.36 & 0.26 & 0.30 & No \\
    & MLP          & 0.75 & 0.54 & 0.62 & Yes \\
    & MLP          & 0.20 & 0.23 & 0.25 & No \\
\midrule
\multirow{6}{*}{Tooling} 
    & BERT         & 0.80 & 0.99 & 0.89 & No \\
    & DistilBERT   & 0.99 & 0.99 & 0.99 & No \\
    & CNN          & 0.88 & 0.91 & 0.89 & Yes \\
    & CNN          & 0.47 & 0.45 & 0.46 & No \\
    & MLP          & 0.64 & 0.68 & 0.65 & Yes \\
    & MLP          & 0.47 & 0.47 & 0.47 & No \\
\midrule
\multirow{6}{*}{Errors} 
    & BERT         & 0.99 & 0.77 & 0.87 & No \\
    & DistilBERT   & 0.99 & 0.94 & 0.97 & No \\
    & CNN          & 0.93 & 0.95 & 0.94 & Yes \\
    & CNN          & 0.77 & 0.77 & 0.77 & No \\
    & MLP          & 0.81 & 0.87 & 0.84 & Yes \\
    & MLP          & 0.68 & 0.90 & 0.78 & No \\
\midrule
\multirow{6}{*}{API Usage} 
    & BERT         & 0.99 & 0.98 & 0.99 & No \\
    & DistilBERT   & 0.99 & 0.98 & 0.99 & No \\
    & CNN          & 0.93 & 0.80 & 0.86 & Yes \\
    & CNN          & 0.29 & 0.22 & 0.20 & No \\
    & MLP          & 0.59 & 0.44 & 0.49 & Yes \\
    & MLP          & 0.28 & 0.24 & 0.27 & No \\
\midrule
\multicolumn{5}{c}{\textbf{Overall Accuracy}} & \\
\midrule
\multicolumn{2}{l}{BERT}           & \multicolumn{3}{c}{95\%} & No \\
\multicolumn{2}{l}{DistilBERT}     & \multicolumn{3}{c}{95\%} & No \\
\multicolumn{2}{l}{CNN }   & \multicolumn{3}{c}{86\%} & Yes \\
\multicolumn{2}{l}{CNN}  & \multicolumn{3}{c}{51\%} & No \\
\multicolumn{2}{l}{MLP}   & \multicolumn{3}{c}{67\%} & Yes \\
\multicolumn{2}{l}{MLP}  & \multicolumn{3}{c}{54\%} & No \\
\hline
\end{tabular}
\end{table*}

\begin{figure}[h]
  \centering
  \includegraphics[width=\linewidth]{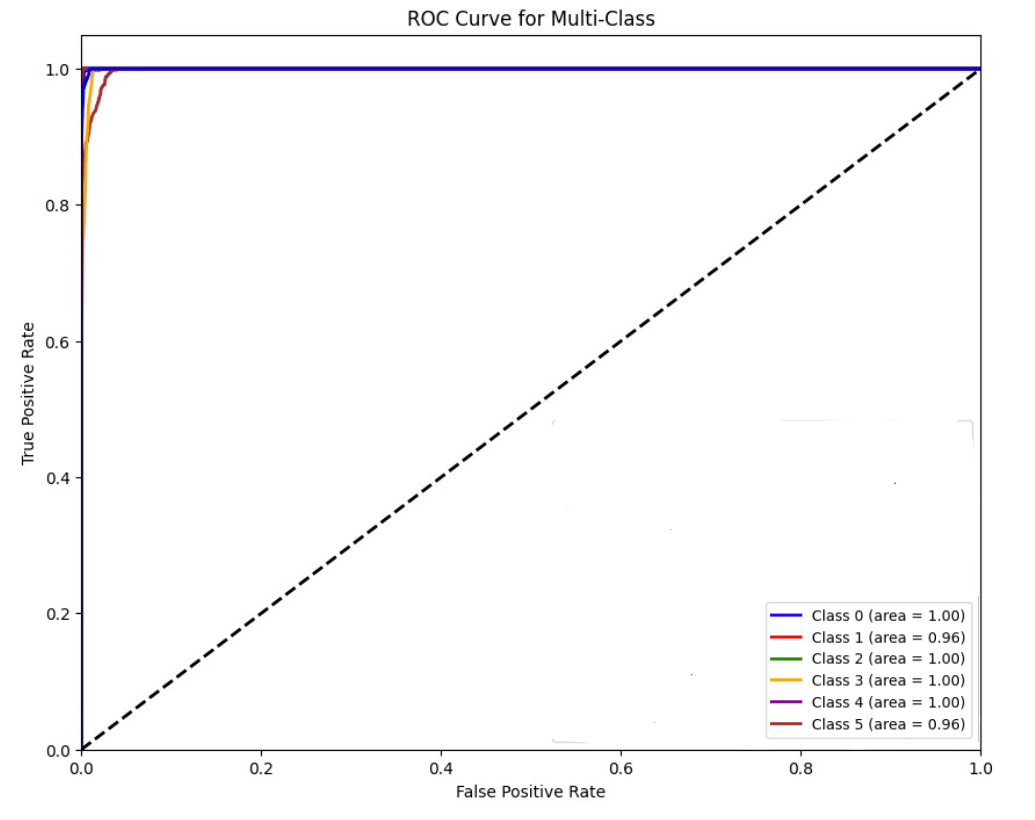}
\caption{ROC curves for BERT}
\label{fig:roc_bert}
\end{figure}

Moreover, we examined the baseline configurations of the top-performing classifiers, BERT and DistilBERT. As a central part of this analysis, we examined the Receiver Operating Characteristic (ROC) curves. ROC curves compare the true positive rate with the false positive rate to determine whether a classifier is sensitive or specific. By examining the accuracy metrics and the ROC curves, an analysis of the BERT model in organizing developer discussions from online distinct thematic categories is conducted. With AUC values near the ideal mark, as depicted in Figure \ref{fig:roc_bert}, each category demonstrated nearly perfect separation. This is evidenced by a high True Positive Rate (TPR) and a low False Positive Rate (FPR), with the curves approaching the upper-left corner of the plot. All three models achieved high macro-average AUC scores, indicating their consistent and precise ability to classify all categories accurately. These metrics highlight the proficiency of the BERT and DistilBERT models in accurately recognizing each discussion category while minimizing misclassifications, thereby confirming their robust performance in multiclass classification

\begin{figure}[h]
  \centering
  \includegraphics[width=\linewidth]{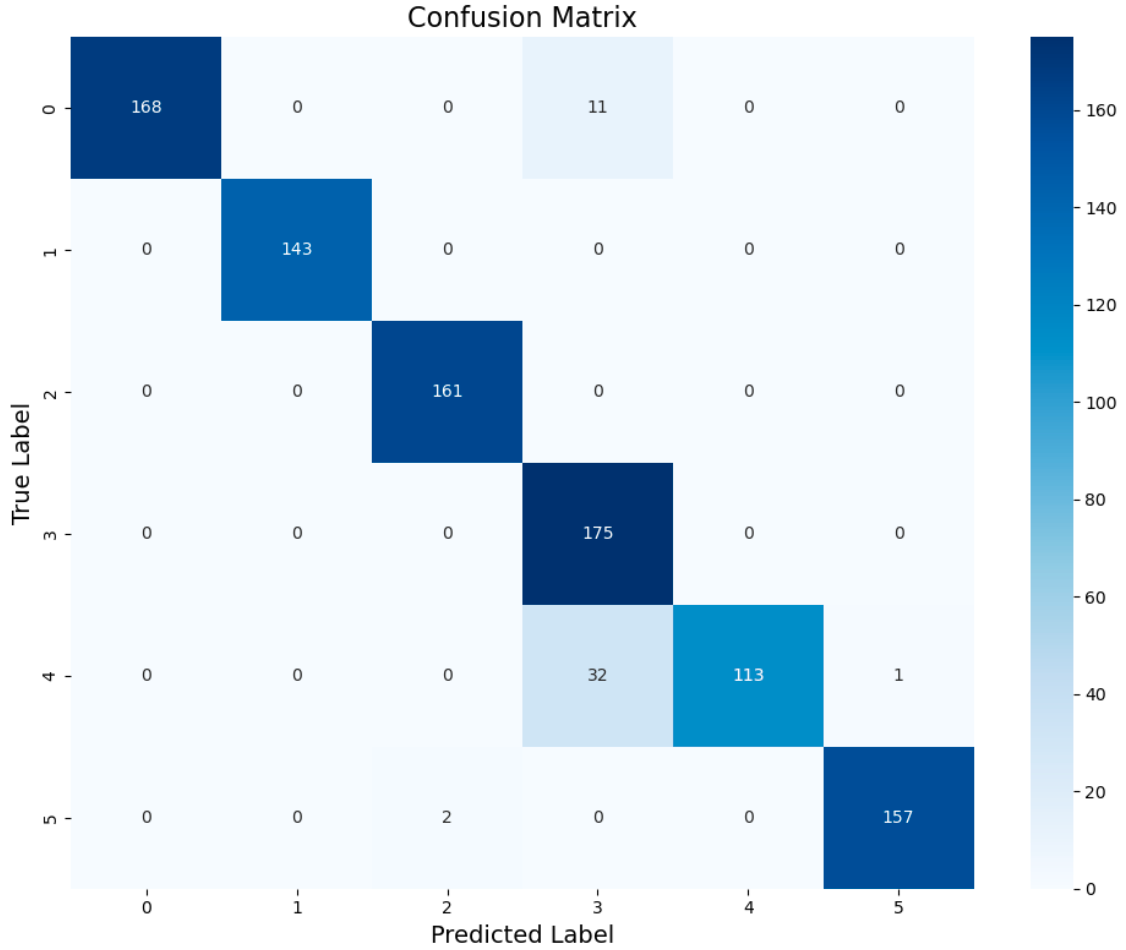}
\caption{Confusion Matrix for BERT (0=Tooling, 1=Conceptual, 2=Errors, 4=API Usage, 5=Learning).}
\label{fig: confusion_bert}
\end{figure}

\begin{figure}[h]
  \centering
  \includegraphics[width=\linewidth]{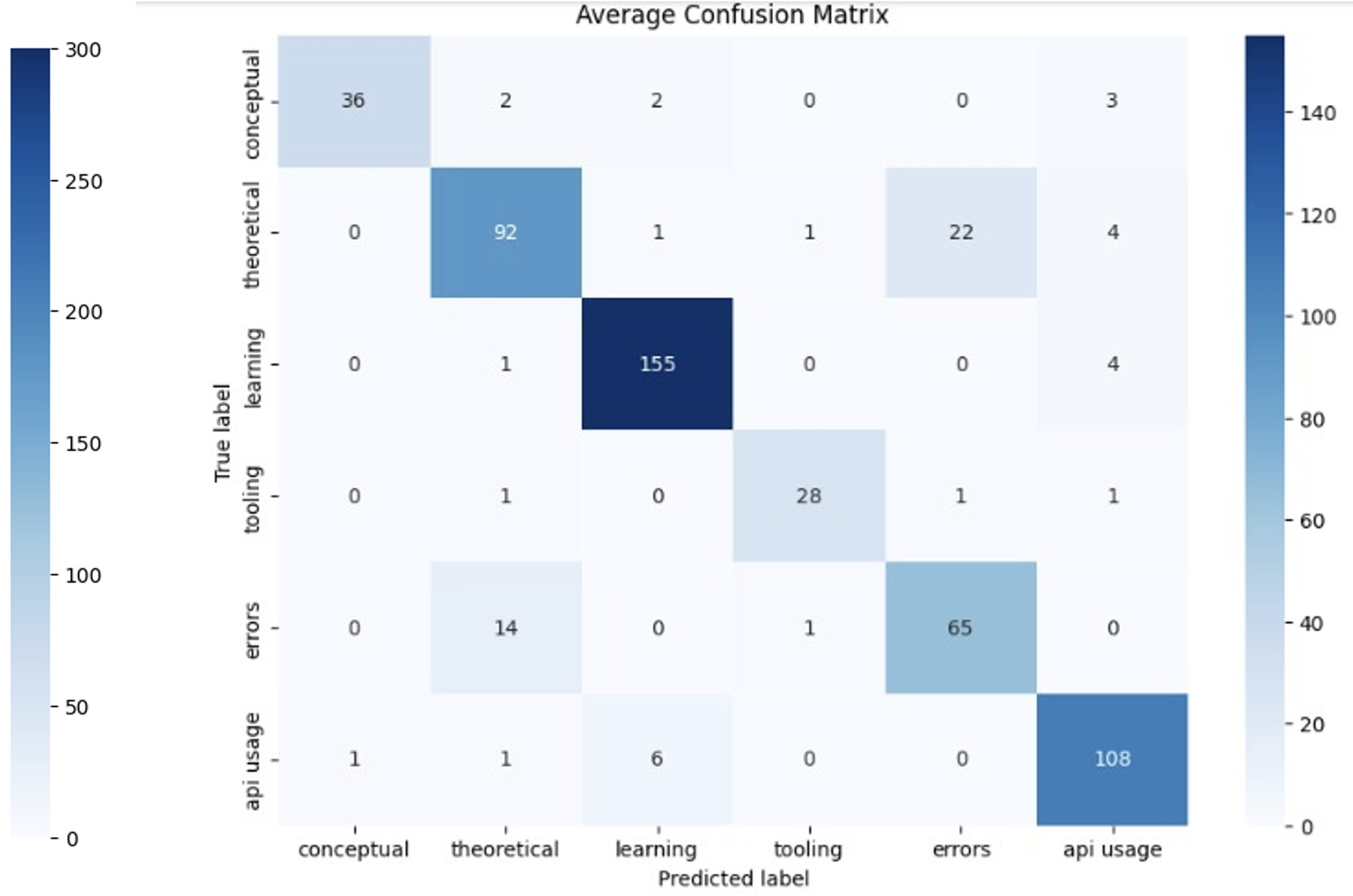}
\caption{Confusion Matrix for CNN model with Data augmentation.}
\label{fig: confusion_CNN_augmentation}
\end{figure}

\begin{figure}[h]
  \centering
  \includegraphics[width=\linewidth]{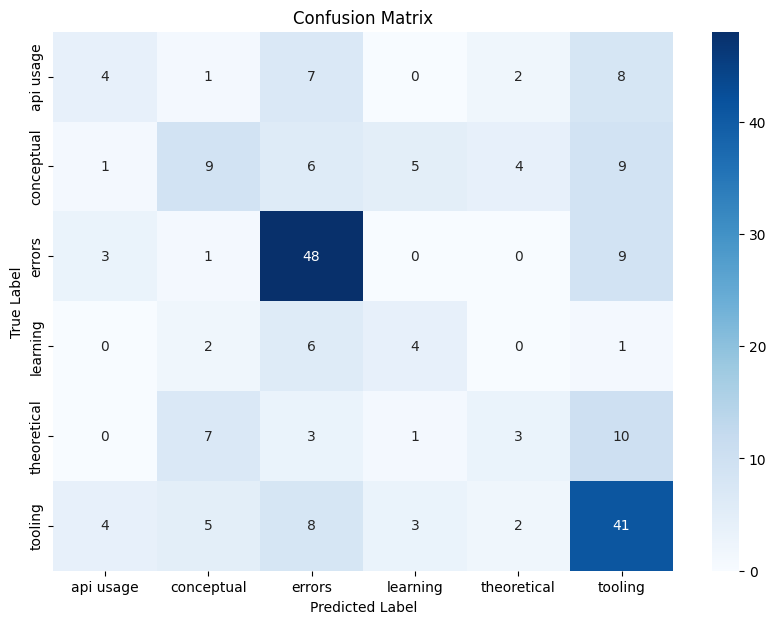}
\caption{Confusion Matrix for CNN model Without Data augmentation .}
\label{fig: confusion_cnn_without_augmentation}
\end{figure}

Furthermore, we compared the performances of the fine-tuned transformer  (BERT) and deep learning (CNN) models by drawing their confusion matrices to identify false positives and negative rates. As shown in Figures \ref{fig: confusion_bert}, \ref{fig: confusion_CNN_augmentation}, and \ref{fig: confusion_cnn_without_augmentation}, the fine-tuned BERT outperforms CNN classifiers both with and without augmentation, achieving low rates of false positives and negatives, resulting in more generalized classification results. In these confusion metrics, the row labels show the actual classes, while the column class labels show the predicted classes for each model. A confusion matrix is used to evaluate the performance of TL classifiers by summarizing the total number of correct and incorrect predictions, broken down by each classification class. In Figure \ref{fig: confusion_bert}, the diagonal values, i.e., 168, 143, etc., demonstrate the true positive values for each QSE category, such as tooling, conceptual, etc., showing how many developers' discussions in the Q\&A forums are identified correctly. When comparing it to the diagonal values of CNN classifiers both with and without augmentation, as shown in Figures  \ref{fig: confusion_CNN_augmentation} and \ref{fig: confusion_cnn_without_augmentation}, respectively. The fine-tuned BERT algorithm performed better and achieved better generalizability. In contrast, the off-diagonal values show False positives and False negatives values, in each confusion matrix, as shown in Figures \ref{fig: confusion_bert}, \ref{fig: confusion_CNN_augmentation}, and \ref{fig: confusion_cnn_without_augmentation}, respectively. When compared, the fine-tuned BERT classifier shows comparatively lower rates of false positive and false negative values. These experimental findings demonstrate that fine-tuned transformer outperforms fine-tuned D\&ML algorithms with relatively smaller and complex datasets, due to its extensive training on large corpora and contextual understanding ability of encoders, which enables them to process complete sentences simultaneously, instead of word by word.

\subsection{Global Interpretability of the BERT Model via SHAP Analysis}
\begin{figure*}[htbp]
\vspace{-10pt}
\centerline{\includegraphics[width=0.9\linewidth ]{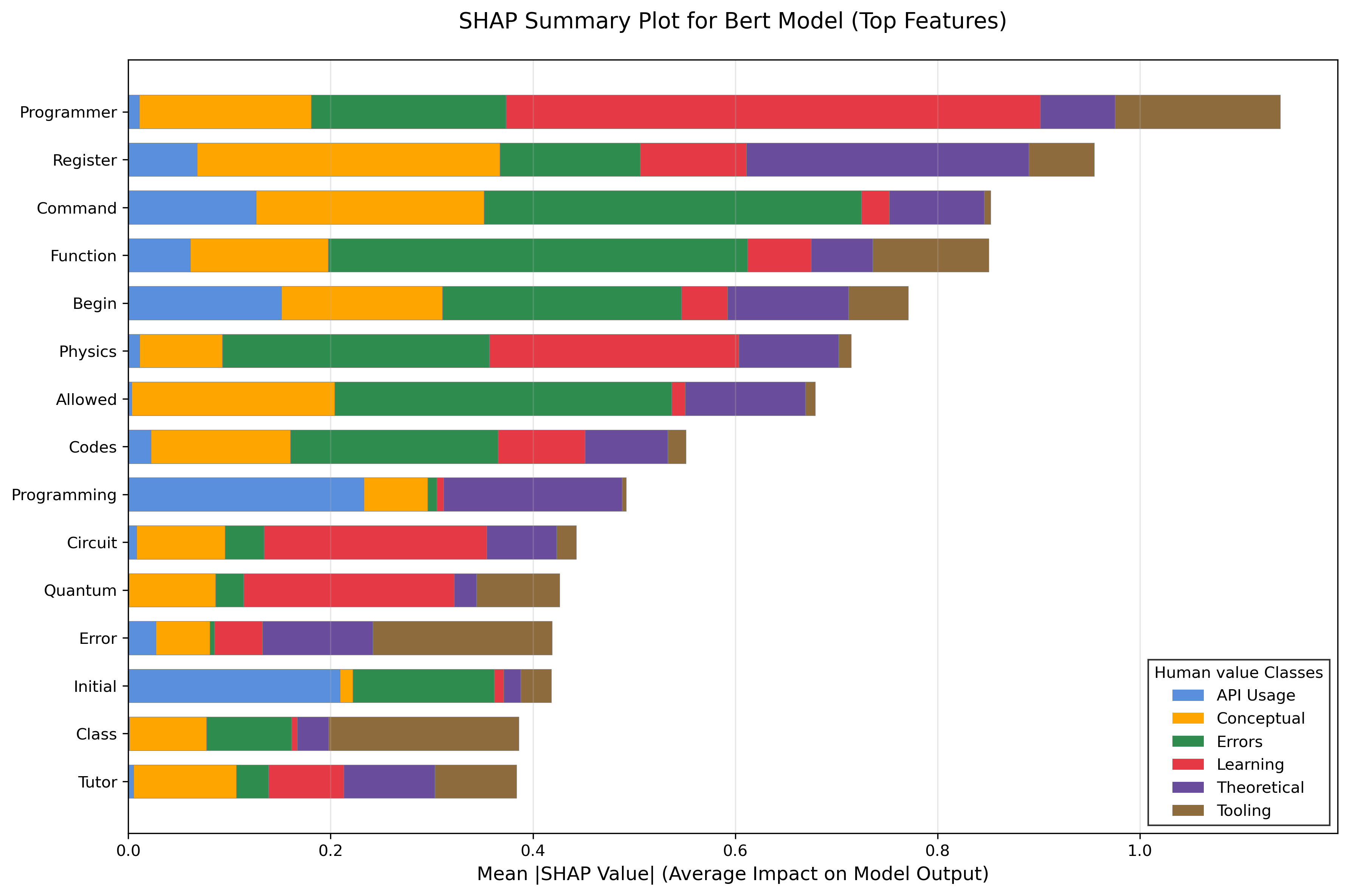}}
\vspace{-10pt}
\caption{SHAP summary Plot for BERT Model.}
\label{fig:shap_bert}
\vspace{-8pt}
\end{figure*}

To enhance transparency in the BERT classifier’s decision-making process, we applied SHAP to identify the most influential textual features across all QSE challenge categories. The resulting SHAP summary plot, shown in Figure~\ref{fig:shap_bert}, ranks the top features by their mean absolute SHAP value, the average impact on model output revealing which terms drive classification decisions. Each horizontal bar represents a key token such as "Programmer", "Register", or "Function", ordered by its overall importance, with color segments indicating the contribution of that feature to specific QSE categories: API Usage (blue), Conceptual (orange), Errors (green), Learning (red), Theoretical (purple), and Tooling (brown).

The highest-impact feature, "Programmer", exhibits a mean |SHAP| value of approximately 1.05, with dominant contributions from \textit{Learning} (red) and \textit{Conceptual} (orange), followed by \textit{Tooling} (brown). This suggests that discussions involving programming roles frequently center on educational challenges or conceptual understanding, particularly in contexts where users require guidance or clarification. Similarly, "Register" has a strong influence (~0.98), primarily linked to \textit{API Usage} (blue) and \textit{Conceptual} (orange), suggesting its relevance in discussions about hardware abstraction layers or low-level programming constructs. "Command" and "Function" also rank highly, with substantial contributions from \textit{Errors} (green) and \textit{API Usage} (blue), reflecting their frequent appearance in error messages or API invocation queries.
\\
Notably, "Physics" shows a balanced mix of \textit{Conceptual} (orange) and \textit{Theoretical} (purple), highlighting the interplay between foundational theory and practical implementation in quantum computing discourse. Terms like "Circuit", "Quantum", and "Error" are strongly associated with \textit{Tooling} (brown) and \textit{Errors} (green), underscoring their role in debugging and simulation workflows. "Initial" and "Class" show significant contributions from \textit{API Usage} (blue) and \textit{Tooling} (brown), indicating their relevance in code initialization and object-oriented design contexts. Overall, this analysis demonstrates that the BERT model leverages both technical terminology and contextual semantics to make nuanced predictions, enabling developers to trace classification decisions back to specific linguistic cues and domain concepts.

\subsection{Local Interpretability via SHAP Force Plot}
\begin{figure*}[htbp]
\centerline{\includegraphics[width=1.0\linewidth]{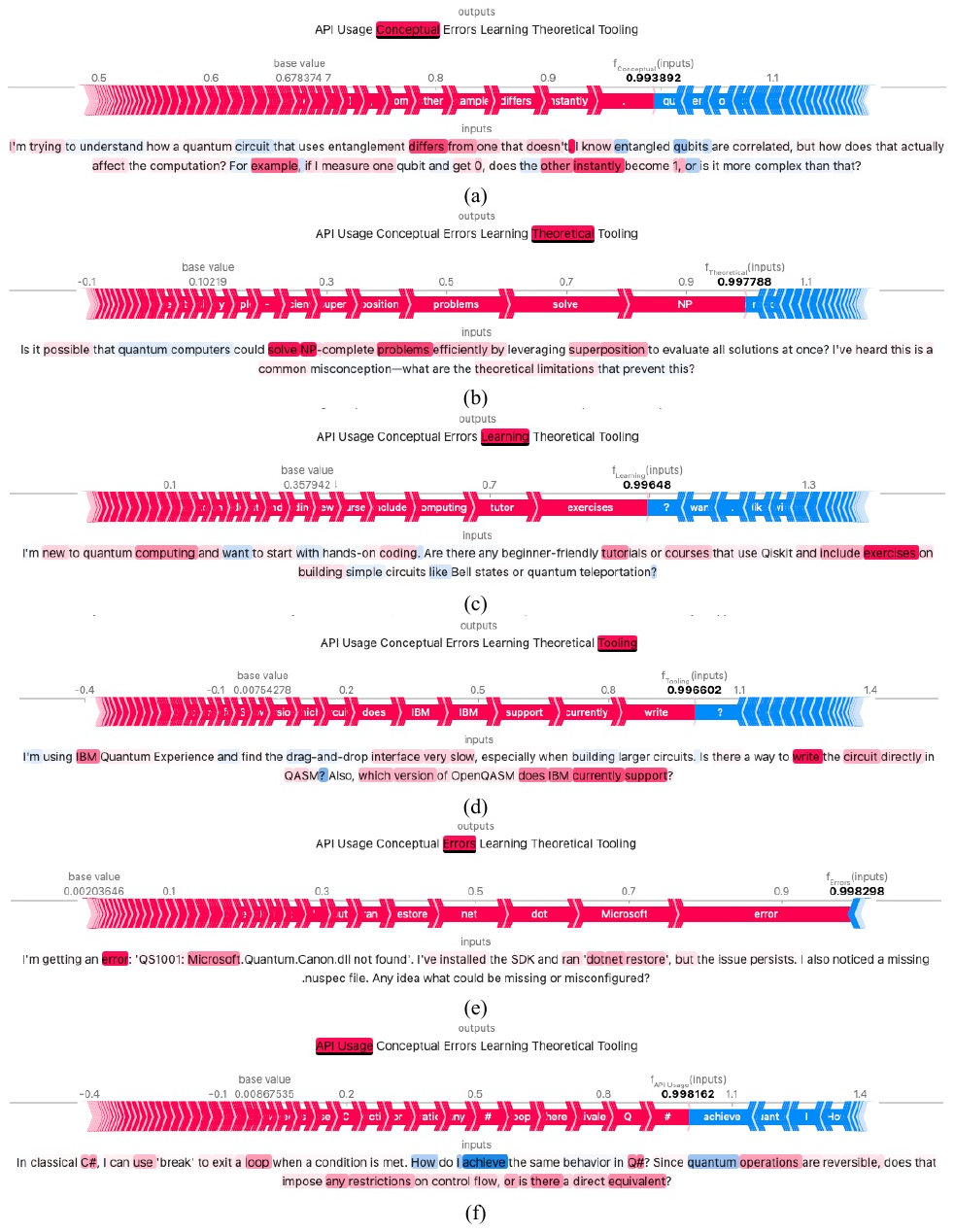}}
\vspace{-12pt} % Reduce space below image
\caption{SHAP Force Plots for BERT Model: (a) \textit{Conceptual}, (b) \textit{Theoretical}, (c) \textit{Learning}, (d) \textit{Tooling}, (e) \textit{Errors}, and (f) \textit{API Usage}.}
\vspace{-10pt} % Reduce space below caption
\label{fig:shap_force_bert}
\end{figure*}

To further elucidate the model's decision-making process at the instance level, we present SHAP force plots for representative examples from each of the six QSE challenge categories. These visualizations break down individual predictions into contributions from specific input tokens, revealing how linguistic cues influence classification outcomes across various domains. As shown in Figure~\ref{fig:shap_force_bert}(a), the model's high confidence in classifying the input as \textit{Conceptual} (0.993892) is driven by key terms such as "example", "differ", "instantly", and "trying" highlighted with red color, which contributed positively in the predication of Conceptual QSE challenge type. Moreover, "qubits", "circuit", and "understand" keywords contribute negatively to the predication of the conceptual QSE challenge and force it onward to other QSE challenges. However, their contribution to the developer feedback is not high. Additionally, the phrase "how does that actually affect the computation?" further reinforces the abstract nature of the query, indicating a focus on fundamental behavior rather than implementation or debugging. This analysis demonstrates that the BERT model effectively identifies theoretical questions by recognizing language centered on quantum mechanics principles, distinguishing them from discussions related to learning, tooling, or errors.

Furthermore in Figure~\ref{fig:shap_force_bert}(b), the model's high confidence in classifying the input as \textit{Theoretical} (0.997788) is driven by key terms such as "NP-complete", "problems", "superposition", and "theoretical limitations", which reflect deep-level inquiry into quantum computing’s computational foundations. The phrase "solve NP-complete problems efficiently" signals a question about algorithmic complexity, while "leveraging superposition" highlights an understanding of quantum mechanics principles. The mention of "common misconception" further reinforces the abstract nature of the query, indicating that the user is engaging with foundational knowledge rather than implementation or debugging. While the "quantum computers" appear in blue adds a negative contribution and skews towards the other QSE challenges. However, their contribution is not extensive. This analysis demonstrates that the BERT model effectively distinguishes theoretical challenges by recognizing language centered on computational models, complexity theory, and scientific boundaries.

In addition Figure~\ref{fig:shap_force_bert}(c), the model's high confidence in classifying the input as \textit{Learning} (0.99648) is driven by terms such as "tutorials", "courses", "exercises", and "computing", shown with red color (the darker, the more positive impact in the prediction) which collectively signal a request for educational resources. While "computing" appears in red, indicating a positive contribution, it is part of the phrase "quantum computing", a domain-specific term often associated with learning pathways. In contrast, "want" appears in blue, reflecting its contextual ambiguity; while commonly used in learning queries, it also appears in requests for tools or fixes, making it less discriminative. This analysis demonstrates that the BERT model distinguishes learning-related discussions by prioritizing domain-specific keywords over generic intent verbs, enabling accurate classification despite linguistic overlap across categories.

Notably, Figure~\ref{fig:shap_force_bert}(d), the model's high confidence in classifying the input as \textit{Tooling} (0.996602) is driven by terms such as "IBM Quantum (with IBM more prominent)", "interface very slow", "QASM", "which version", and "write", which collectively signal a discussion about software tool usability and workflow efficiency. The phrase "interface very slow" indicates performance concerns, pointing to a possible tool. In contrast, words such as "?", "I'm" and "drag-and" contribute relatively negatively in the predication of the tool QSE category, but with less intensity. This analysis demonstrates that the BERT model effectively identifies tooling challenges by recognizing language centered on software interfaces, performance limitations, and syntax-specific workflows.

Conversely, Figure~\ref{fig:shap_force_bert}(e), the model's high confidence in classifying the input as \textit{Errors} (0.998298) is driven by explicit technical indicators such as "error", "dotnetrestore", and "Microsoft", which collectively signal a software fault or dependency issue. The phrase "issue persists" reinforces the ongoing nature of the problem, while "dotnet restore" indicates an attempted resolution, confirming this as a troubleshooting scenario. Although "installed" and "SDK" appear in red, their context here supports error diagnosis rather than successful setup. This analysis demonstrates that the BERT model effectively identifies error-related discussions by recognizing language centered on specific error codes, missing files, and failed recovery attempts, distinguishing them from conceptual or learning queries.

Finally, Figure~\ref{fig:shap_force_bert}(f), the model's high confidence in classifying the input as \textit{API Usage} (0.998162) is driven by technical terms such as "equivalent", "loop", "any restriction", "Q\#", and "operations", which reflect a query about programming semantics in quantum languages. The words "how", "quantum", and "achieve", as shown in blue, contribute negatively to the prediction of API Usage. However, the intensity of these words is lower when deciding on classifying the developers' discussion into the more suitable QSE category. This analysis demonstrates that the BERT model effectively identifies API usage challenges by recognizing language centered on syntax, control structures, and functional equivalence between classical and quantum paradigms.

\subsection{Comparative Study with the State of the Art}
This section compares the proposed approach with existing methods in terms of scope, performance, and generalization. An overview of the comparison is shown in Table 9. Beyer et al.\cite{beyer2020kind} use ML algorithms, including RF and SVM, to categorize Android-related discussions on SO. Their approach achieved an average accuracy of 90\%, and it is limited to classifying developers' posts into Android-related challenges. In contrast, the proposed approach utilizes state-of-the-art transfer learning algorithms to contextually analyze developer discussions, achieving an overall accuracy of 95\% and lower values of false positives and negatives. Yousuf and Sofi \cite{yousuf2025bug} proposed a rule-based approach to classify frequently occurring issues in quantum Qiskit repositories, achieving an accuracy of 85.2\%. The proposed approach complements the El-Aoun et al. approach \cite{li2021understanding} by focusing on QSE from various Q\&A platforms. However, we are classifying developers' discussions into frequently occurring challenges, whereas their approach identifies QSE challenges using topic modeling by analyzing developers' feedback. We argue that the proposed approach can help improve developers' experiences in quickly and better identifying related information on these Q\&A forums by categorizing it based on the frequently occurring challenges, in addition to the existing tags. In addition, Aktar et al. \cite{aktar2025architecture} conducted a qualitative analysis of discussions about quantum software development on Stack Exchange forums and GitHub issues. By examining 385 discussions/topics, they identified common challenges and solutions related to architecture decisions in quantum systems. The study categorized the decisions into implementation and technology-related issues, providing valuable insights into the linguistic patterns used by developers when discussing quantum software architectures. Furthermore, Upadhyay et al. \cite{upadhyay2025analyzing} conducted a mining analysis on more than 21,000 quantum computing repositories, with over 1.2 million commits from more than 10,000 developers. The study aimed to uncover evolution and maintenance practices in quantum software development, highlighting key trends in programming languages, framework adoption, and developer growth. This large-scale analysis provides comprehensive insights into software engineering challenges and offers interesting insights into improving quantum development workflows, documentation, and tooling. In addition to our previously published work \cite{husain2025exploring}, this study demonstrates comparatively better precision, recall, F-measure, and overall accuracy by employing fine-tuned transformer algorithms, including BERT, RoBERTa, and DistilBERT, by contextually analyzing developers' discussions from various Q\&A platforms and classifying them into frequently occurring QSE challenges with lower false negatives and positives rates. Overall, the proposed approach achieves 95\% accuracy, outperforming our previous work \cite{husain2025exploring} by approximately 6\% (with data augmentation) and 41\% without data augmentation. Additionally, the proposed approach enhances explainability by experimenting with SHAP, which identifies the frequent features used by the BERT classifier when classifying a particular developer discussion as a QSE challenge. Moreover, the proposed approach demonstrates why a particular developer's feedback is classified as a QSE challenge, along with its overall confidence rate. These insights can help quantum practitioners categorise Q\&A posts based on the frequently identified challenges, providing easy access to related posts and solutions from other developers as an alternative option to the existing tags-based search.

\begin{table*}[t]
\centering
\footnotesize
\caption{Comparative Study with the State of the Art}
\label{tab:comparative_study}
\begin{tabular}{p{2.0cm}p{3.2cm}p{4.5cm}p{2.5cm}c}
\hline
\textbf{Reference Papers} & \textbf{Method Used} & \textbf{Highest Precision, Accuracy, Recall} & \textbf{Dataset Details} & \textbf{Data Size} \\
\hline

\cite{beyer2020kind}& 
Random Forest, SVM & 
Precision 90\%, Accuracy 90\%, Recall 90\% & 
Android-related SO posts & 
1,000 posts \\
\cite{yousuf2025bug} & 
Rule-based classification & 
Accuracy 85.2\% & 
Qiskit issues (36 repos) & 
12,910 issues (sampled 4,984) \\

\cite{li2021understanding} & 
Automated Topic Modeling & 
N/A & 
Stack Exchange forums, GitHub issues & 
1,755 issues/posts \\
\cite{aktar2025architecture} & 
Qualitative analysis & 
N/A & 
Stack Exchange forums, GitHub issues & 
385 discussions/issues \\
\cite{upadhyay2025analyzing} & 
Mining analysis & 
N/A & 
21,000+ repositories & 
1.2 million+ commits \\

\textbf{Our Study} & 
\textbf{BERT, RoBERTa, and DistilBERT} & 
\textbf{Accuracy 95\%} & 
\textbf{Multiple Q\&A forums (SO, QCSE, CSSE, AISE)} & 
\textbf{2,829 discussions} \\
\hline
\end{tabular}
\end{table*}

\section{Discussions} 
The insights obtained from developer interactions on diverse social media platforms are powerful tools for software developers and vendors. This section will delve into our research findings, as elaborated below.

\subsection{ Rationale for Analyzing QSE-related Developers' Discussion}
The recent interest, research findings, and success of quantum computing in various industries have paved the way for the development of quantum software. However, currently there is a lack of software development practices and methodologies for quantum software development \cite{murillo2025quantum}. Also, adopting classical software practices for quantum software development is challenging due to differences in control flow, debugging complexities, computation models, and resource limitations. To address these challenges, researchers started exploring Quantum Software Engineering (QSE) to develop and propose quantum software-specific development methodologies, tools, and practices. It is evident in the literature that Software and quantum developers frequently refer to developers' Q\&A forums to seek software solutions for development-related challenges \cite{ahmad2018survey, li2021understanding}. For this purpose, we analyzed developers' discussions collected from various Q\&A platforms to first annotate them with frequently occurring QSE challenges with humans and ChatGPT and then use this dataset to train a TL algorithm to automate the process. The proposed approach can help software developers quickly and accurately identify QSE-related information if the current Q\&A interfaces are modified. For example, an additional menu can be added to display and categorize developers' discussions based on various QSE categories.  

\begin{table*}[t]
\begin{promptbox}[Key Research Findings 1: ]\label{box:1}
QSE is an emerging area that poses several challenges in realizing its potential for quantum software development. New software quantum development methodologies, tools, and practices are required to overcome these challenges. Mining developers' discussions from Q\&A platforms can provide opportunities for software quantum researchers, developers, and vendors to pinpoint frequently occurring challenges and propose remedies for them.
\end{promptbox}
\end{table*}

\subsection{Implications of Research Findings for Software Quantum Developers and Vendors}
Software quantum developers can leverage research findings in several directions to improve the existing quantum development process and enhance their skills by contributing more effectively to the quantum life cycle. For example, since quantum errors (28\%) have been identified as the most frequently occurring challenge using the content analysis approach, the quantum developers should focus on how to debug and troubleshoot quantum software. This also provides an opportunity for quantum vendors to develop tools that could help in diagnosing and identifying quantum errors in the code. Similarly, quantum developers can allocate time to get expertise in understanding basic quantum mechanisms, such as entanglement, quantum gates, and superposition, as conceptual challenges were reported the second most frequently in the Q\&A platforms. Gaining expertise in this area will enable quantum developers to solve quantum problems more efficiently and enhance software design. These findings encourage software developers to arrange frequent hands-on workshops, tutorials, and lectures on quantum computing to familiarize developers with the emerging quantum concepts and development. Discussions on tooling inefficiencies highlight the gaps between resources and developer needs, suggesting opportunities for more intuitive tools, especially for quantum computing projects. Quantum developers should familiarize themselves with the limitations of existing tools and learn about best practices for implementing and using these tools to achieve improved outputs. At the same time, more fine-grained analysis should provide insights for the quantum vendors to identify areas where quantum developers have confronted challenges to develop more developer-friendly tools. Furthermore, API usage indicates that better documentation and standards can enhance productivity in the field. Quantum developers can take advantage of open source projects and community-driven examples to understand quantum APIs' pitfalls. Quantum vendors and researchers should work on improving the documentation and toy examples to improve the usability of existing quantum API's. Additionally, quantum developers should align with the latest research findings to enhance their theoretical understanding and identify ambiguous quantum areas. Quantum developers could take this opportunity to develop quantum-specific research and development communities to reduce the gap between the theoretical and practical implementation of quantum concepts. Finally, considering the growing interest and success of quantum computing, existing computer science, engineering, and software curricula should incorporate quantum-specific courses.

\begin{table*}[t]
\begin{promptbox}[Key Research Findings 2: ]\label{box:2}
The research findings could benefit quantum developers by mastering conceptual understanding, error handling, and tooling to overcome the most frequently occurring challenges. It would help developers improve their problem-solving skills, optimize tool usage, and effectively contribute to the quantum development life cycle. Similarly, quantum vendors develop tools for debugging errors, improving documentation, and arranging regular workshops and training to enhance software productivity. 
\end{promptbox}
\end{table*}

\subsection{Implications of Transformer-Based Learning for QSE}
The superior performance of TL models, BERT, DistilBERT, and RoBERTa, over traditional DL architectures such as FNN, CNN, and RNN highlights a fundamental shift in how we approach text classification in emerging software engineering domains like QSE. Unlike DL models that rely on shallow feature extraction and local syntactic patterns, TL models leverage large-scale pre-training on general language corpora and fine-tuning on domain-specific data, enabling them to capture deep semantic relationships, contextual nuances, and long-range dependencies in developer discourse. This architectural advantage translates into significantly higher accuracy (95\% vs. 86\% for CNN (with data augmentation)) and more robust generalization across diverse challenge categories such as \textit{Conceptual}, \textit{Theoretical}, and \textit{API Usage}. The improvements of TL algorithms become more prominent (95\% vs. 51\% for CNN (without data augmentation)) when compared to D\&ML algorithms while processing developers' discussion without data augmentation. Additionally, fine-tuned transformer algorithms result in more reliable performances by minimizing the false positives and negatives rates.
\\
A key reason for this performance gap lies in the nature of quantum programming discussions, which are often abstract, jargon-rich, and context-dependent. Terms like "entanglement", "superposition", or "Q\#" carry specific meanings that vary subtly based on surrounding context—a nuance that DL models struggle to grasp due to limited contextual receptive fields. In contrast, TL models use self-attention mechanisms to dynamically weigh the importance of every word in a sentence, allowing them to distinguish between a query about theoretical complexity ("Can quantum computers solve NP-complete problems?") and one about practical implementation ("How do I write a loop in Q\#?"). This sensitivity to semantic context makes TL models particularly well-suited for classifying unstructured, natural language content from forums where developers mix technical detail with personal experience and conceptual uncertainty.

Beyond accuracy, the success of TL models has broader implications for knowledge management and tool support in nascent software domains. Their ability to learn from relatively small, annotated datasets makes them viable even in low-resource settings like QSE, where labeled data is scarce. Furthermore, their compatibility with standard fine-tuning pipelines enables rapid adaptation to new platforms, APIs, or communities. As quantum software evolves, these models can be continuously retrained to reflect new challenges, making them ideal candidates for building intelligent assistants, automated documentation systems, and real-time support tools. The demonstrated effectiveness of TL models in this work sets a precedent for their adoption in other complex, language-intensive domains of software engineering, including AI/ML development, blockchain programming, and embedded systems.

\begin{table*}[t]
\begin{promptbox}[Key Research Findings 3: ]\label{box:3}
Fine-tuned transformer algorithm outperforms fine-tuned D\&ML algorithms both with augmentation and without to classify developers' discussions into various frequently occurring challenges. Additionally, the research findings indicate that fine-tuned transformers yield low rates of false positives and negatives. 
\end{promptbox}
\end{table*}
\subsection{Implications of SHAP for Transformer-Based Models in QSE}
\label{subsec:shap_implications}
The integration of SHAP with transformer-based models like BERT offers profound practical benefits for quantum vendors and developers. By making the model’s decision-making transparent, SHAP enables developers and tool designers to trust and validate automated classification systems used to analyze forum discussions. For instance, when a post is classified as \textit{API Usage} based on terms like "loop", "control flow", and "Q\#", SHAP confirms that the decision is grounded in meaningful, contextually relevant programming constructs, not spurious correlations or noise. This level of justification is crucial for developing intelligent support tools, such as recommendation engines that suggest relevant documentation or code examples based on the detected challenge type. Moreover, project maintainers can use SHAP-generated insights to identify recurring pain points, such as confusion around quantum control structures, and prioritize improvements in language design, error messages, or API documentation. In educational contexts, instructors and curriculum designers can leverage these explanations to refine teaching materials by focusing on the most frequently misunderstood concepts, such as entanglement or superposition, as reflected in the influential terms surfaced by SHAP.

Beyond immediate usability, SHAP empowers researchers and practitioners to perform model-driven knowledge discovery in emerging domains where formal documentation is sparse. The ability to see which words drive predictions, like "QS1001" for \textit{Errors} or "drag-and-drop interface" for \textit{Tooling}, allows teams to detect evolving challenges and adapt their tooling proactively. For example, if SHAP consistently highlights dependency-related terms in error reports, it signals a need for better package management tooling in quantum SDKs. Similarly, the negative contribution of "classical C\#" in an \textit{API Usage} query reveals how developers frame their questions through contrast, suggesting that future tools should support comparative reasoning (e.g., “How is Q\# different from C\# for loops?”). For researchers, this interpretability layer transforms BERT from a classification engine into a \textit{diagnostic instrument} for studying developer cognition, communication patterns, and learning trajectories. As quantum software matures, such explainable AI systems will not only support individual developers but also inform community-wide knowledge curation, making SHAP an essential component of any data-driven QSE research pipeline.
\begin{table*}[t]
\begin{promptbox}[Key Research Findings 4: ]\label{box:1}
Integrating an Explainable AI approach, such as SHAP, with the fine-tuned transformer-based approach enhances its operability and makes the decision-making process more transparent. Quantum vendors can utilize the research findings to enhance the performance of existing classification algorithms by incorporating these keywords as features. 
\end{promptbox}
\end{table*}
\subsection{Threats to Validity}
While processing and analyzing developer discussions, we encountered several potential threats to the validity of the proposed methodology and findings. Acknowledging these challenges is crucial to ensuring the credibility and reliability of the research outcomes.
\subsubsection{Researcher Bias}
In the proposed study, the potential for researcher bias is significant, as the first, second, and third authors, along with an AI model such as ChatGPT, labeled and analyzed developer discussions to maintain neutrality and objectivity. However, our inherent attitudes and previous experiences subtly affected the analysis. This bias could skew the data's performance or emphasize specific findings. Acknowledging the presence of such biases is crucial to ensuring the integrity and reliability of our research findings.
To mitigate this threat, we employed a human-AI collaborative annotation process, where disagreements between human coders and ChatGPT were resolved through negotiation and justification. This hybrid approach improved annotation consistency, achieving a Cohen’s kappa of 0.46 indicating substantial agreement and reducing individual subjectivity in labeling.
\subsubsection{Data Preprocessing Challenges }
Data preprocessing is a critical step in the proposed approach, fraught with challenges, including addressing noisy data, removing biases, and resolving semantic opacity in developer discussions. These issues can significantly impact the quality and interpretability of data, potentially affecting subsequent analyses. Therefore, we must implement rich data preprocessing techniques to mitigate these issues, thereby enhancing the validity of our findings by ensuring that the data fed into our analytical models is as accurate and clean as possible.
\subsubsection{Algorithm Selection and Fine Tuning} 
The selection and fine-tuning of D\&ML algorithms to examine developer discussions are fraught with several challenges. The proposed research's effectiveness and applicability depend heavily on selecting appropriate algorithms and optimizing them to handle the specific nuances of the QSE data. This process requires deep expertise and careful experimentation, as the selection of an imperfect algorithm or insufficient tuning can lead to misleading outcomes or decreased model effectiveness. We addressed this threat by conducting extensive comparative evaluation across multiple architectures, including FNN, CNN, RNN, and TL models, and using cross-validation to ensure stable performance estimation. Furthermore, the integration of SHAP provided post-hoc interpretability, allowing us to validate whether model decisions were driven by meaningful features rather than artifacts of overfitting or poor tuning.
\subsubsection{Data Quality Concerns}
Data quality is paramount for ensuring the reliability of the machine-learning models used in the study. High-quality data training and evaluation are essential for producing valid and generalizable results. Data noise, inherent biases, and semantic ambiguities within developer discussions must be addressed rigorously. These factors must be considered to maintain the trustworthiness of the research outputs, as the models may learn from flawed data, leading to unreliable predictions and insights. Our dataset was curated through a rigorous grounded theory and content analysis approach, resulting in 2,829 labeled instances. Conflicts in initial annotations were systematically resolved using ChatGPT-generated rationales, leading to a conflict-free final dataset. This iterative validation process significantly enhanced data quality and reduced the risk of propagating labeling errors into model training.
\subsubsection{Ever-Evolving Landscape of Online Forums}
The vibrant and ever-changing landscape of online forums and developer communities presents additional challenges for maintaining the scalability and generalizability of the proposed approach. These platforms have evolved rapidly with new models and continually emerging topics. Thus, the proposed methodology requires ongoing adjustments and updates to remain relevant and effective in capturing the current state of developer discussions across various forums. To address this, the proposed framework is designed to be extensible, supporting periodic retraining with newly collected data. Moreover, the use of pre-trained language models like BERT improves adaptability, as they can be efficiently fine-tuned on updated corpora. Future work will explore automated drift detection mechanisms to trigger model updates when shifts in discussion patterns are identified.
\section{Conclusions and future work}
This study presents a TL–based approach to classifying developer challenges in QSE-related frequently occurring challenges, achieving 95\% accuracy through fine-tuned TL-based models, BERT, and DistilBERT, outperforming traditional DML methods. By analyzing discussions from multiple Q\&A forums and employing SHAP for explainability, the model not only delivers high performance, but also provides interpretable insights into feature contributions, revealing key linguistic patterns associated with challenge categories such as Conceptual, Learning, Error, ApI Usage, theoretical, and Tooling. The integration of SHAP provides both global and local interpretability, revealing how specific terms influence model predictions. This work advances the automation of knowledge extraction in QSE, offering a scalable, transparent framework to support the growing quantum development community.

Future studies will focus on key areas to expand and enhance the capabilities of the automated categorization model. We plan to explore other forums and discussion platforms popular in the QSE community to collect more data. We aim to provide a more intuitive interface and toolchain that classifies developers' discussions into frequently occurring challenges on the run. These efforts aim to develop robust collaborative relationships with field professionals, maintaining the model's relevance and effectiveness in addressing practical barriers in QSE. Moreover, we plan to extend the proposed approach to other emerging technologies' discussions on Q\&A forums, such as blockchain, Large language models, deep learning, and transfer learning. Additionally, we aim to develop a toolset for the proposed approach by implementing the best-performing D\&ML classifiers and demonstrating their applicability for quantum software developers.

\textbf{Authorship contributions:} N.D.K., J.A.K., and M.H. developed the method, detailed investigation, and manuscript writing and curated the research data set; M.S.K., A.A.K., S.H., and M.A.A. revised the methodology and supervised and revised the manuscript writing (revised draft). All authors have read and agreed to the published version of the manuscript.

\textbf{Funding:} This work has been supported by the following projects: 
\textit{Securing the Quantum Software Stack (SeQuSoS), project number 24304955}, funded by Business Finland; and 
\textit{Classi\textbar Q$\rangle$, project number 24304728121}, funded jointly by the City of Oulu and the University of Oulu, Finland.

\bibliographystyle{fcs}
\bibliography{ref}

\begin{biography}{nekdil.JPG}
NEK DIL KHAN received his B.Sc. degree in software engineering from the University of Science and Technology Bannu, Khyber Pakhtunkhwa,
Pakistan. He continued to pursue his passion and earned his master’s degree in software engineering from Northeastern University, Shenyang, Liaoning, China. He is pursuing a Ph.D. degree in software engineering at the Beijing University of Technology, Beijing, China. He explores a broad range of topics in the fields of software engineering and computer science. These include natural language processing, software requirements engineering, debugging, information extraction, big data, data mining, human–computer interaction, game theory, sarcasm detection, CrowdRE, argumentation and argument mining, mining software repositories, human values in software, quantum software engineering, feedback analysis, empirical software engineering, sentiment and opinion mining, requirements prioritization, mining fake reviews, and health analytics.
\end{biography}

\begin{biography}{javed.jpg}
JAVED ALI KHAN received the Ph.D. degree in software engineering from Tsinghua University (QS ranked 17th), Beijing, China. He is currently working as a Senior Lecturer with the Foundation of Software Engineering (FSE) Group, Department of Computer Science, University of Hertfordshire, U.K., teaching core Software Engineering courses. Previously, he worked as the
Assistant Professor cum Chairperson with the Department of Software Engineering, University
of Science and Technology Bannu, Pakistan. His areas of research interest include software engineering,
requirements engineering, CrowdRE, mining software repositories, human values in Software, quantum software
engineering, empirical software engineering, NLP, and health informatics. Javed has published more than 90 peer-reviewed conference and journal papers.
\end{biography}

\begin{biography}{aaa.jpeg}
MOBASHIR HUSAIN received his B.Sc. in software engineering from the University of Science and Technology, Bannu, Pakistan, in 2021, and his M.S. in computer software engineering from the University of Engineering and Technology, Mardan, Pakistan, in 2024. He worked as a Lab Engineer at UST Bannu and as an IT Officer at Alkhidmat Hospital, Bannu. Since September 2025, he has been a Lecturer at Sarhad University of Science and Information Technology (SUIT), Peshawar, Pakistan. His research interests include software engineering, quantum software engineering, AI, software requirements engineering, human-computer interaction, software systems management, software analytics, NLP, big data, sentiment mining, and AI-driven software engineering.
\end{biography}

\begin{biography}{sohail.jpeg}
\raggedright
Muhammad Sohail Khan is an Associate Professor at the Department of Computer Software Engineering, University of Engineering \& Technology Mardan, Pakistan. He received his M.S. degree from Computer Software Engineering Department, University of Engineering and Technology Peshawar in 2012 and completed his Ph.D. degree from Jeju National University South Korea in 2016. He had been a part of the software development industry in Pakistan as a designer and developer. The major focus of his work is the investigation and application of alternate programming strategies to enable the involvement of masses in Internet-of-Things application design and development. His research interests include IoT, Blockchain, Cloud computing, Machine Learning, End-User Programming, Human-Computer Interactions and Empirical Software Engineering.
\end{biography}

\begin{biography}{Arif.jpg}
ARIF ALI KHAN received the Ph.D. degree in Software Engineering from City University of Hong Kong. He is currently working as an Associate Professor in the M3S Empirical Software Engineering Research Unit at the University of Oulu, Finland. Previously, he held academic positions at the University of Jyväskylä, Finland, and Nanjing University of Aeronautics and Astronautics, China. His research interests include quantum software engineering, empirical software engineering, software process improvement, and evidence-based software engineering. He has published more than 100 peer-reviewed articles and serves as an editor and reviewer for leading software engineering venues. Further details are available at: \url{https://www.oulu.fi/en/researchers/arif-ali-khan}.
\end{biography}

\begin{biography}{muhammad.jpeg}
Muhammad Azeem Akbar (PhD) is an Associate Professor and Adjunct Professor (Docent) of Software Process
Improvement and Management in the Software Engineering Department at LUT University, Finland. He earned his Ph.D.
in Software Engineering from Chongqing University, China, in January 2019, where he focused on requirements change
impact analysis and management in global software development environments. Following his doctorate, he completed a
postdoctoral fellowship at the Nanjing University of Aeronautics and Astronautics (July 2019 – September 2020),
contributing to the “Automated Code Recommendation System” project. From September 2020 to August 2021, he served
as a Senior Researcher at LERO – The Irish Software Research Centre, University of Limerick, where he worked on
Huawei’s “Software Trustworthiness” initiative. He joined LUT University as a Postdoctoral Researcher in September
2021 and was promoted to Associate Professor in September 2023. He has published over 130 research articles in well-
reputed journals and international conferences. He taught and designed several core and advanced courses in software
engineering and has been recognized with excellence in teaching, excellence in instructional technology, and excellence in
academic advising awards. His multidisciplinary research integrates software engineering with LLMs, data analytics,
artificial intelligence, and natural language processing to address real-world challenges in modern software systems.
\end{biography}
\begin{biography}{Picture1.jpg}
I am currently working as an Assistant Professor at the Computer Science and Software Engineering Department, School of Penn State University Behrend, Erie, PA, USA. Before Joining PSU, I have worked as a Pro temp (NTTF) at the Computer and Information Science Department, University of Oregon, Eugene, Oregon, USA. Prior to this, I was working as an Assistant Professor in COMSATS University, Islamabad. I have worked as course chair and acting Director Administration at Namal College, Mianwali an affiliated College of University of Bradford, UK. Moreover, I have worked as a faculty member of University of Peshawar, Center of Applied Epistemics, and CECOS University. I have worked as programmer with SRDC, British Council, Peshawar, Pakistan. I am a research collaborator at SERC, MID Lab, Smart Intelligence team, SERC-KUST, SE-CU (HK) team, SE-NUAA (CN) from September 2014 to date. 

My current research focus is to explore the software engineering practices for High Performance Computing (HPC) software packages and applications to improve their process and benchmark their quality levels, to automate the document generation for open-source software, and to explore the implications of Machine learning, Software fairness analysis, Artificial Intelligence (AI), Big data analysis and soft computing technique in analysis and design of software.
\end{biography}
\end{document}